\documentclass[aip,jcp,reprint,amsmath,amssymb,longbibliography]{revtex4-2}
\usepackage{graphicx,dcolumn,bm,booktabs,xcolor}
\usepackage{xr-hyper}
\usepackage{hyperref}
\graphicspath{{figures/}}

\newcommand{\rnew}[1]{#1}
\newcommand{\bP}{\beta P\sigma^3}
\newcommand{\rev}[1]{#1}
\newcommand{\giuseppe}[1]{#1}
\newcommand{\santi}[1]{#1}
\begin{document}
\title[When soft meets hard: pseudo-hard-sphere non-additive mixtures]{Pseudo-hard-sphere description of symmetric non-additive hard-sphere mixtures: structure, dynamics, and glassy states}
\author{Enrique Lomba}
\email{enrique.lomba@csic.es}
\affiliation{Instituto de Qu\'{\i}mica F\'{\i}sica Blas Cabrera, CSIC, Serrano 119, E-28006 Madrid, Spain}
\author{Santi Prestipino}
\affiliation{Dipartimento di Scienze Matematiche e Informatiche, Scienze Fisiche e Scienze della Terra, Universit\`a degli Studi di Messina, viale F. Stagno d'Alcontres 31, I-98166 Messina, Italy}
\affiliation{Institute for Chemical-Physical Processes, National Research Council (CNR-IPCF), Viale F. Stagno d'Alcontres 37, I-98158 Messina, Italy}
\author{Giuseppe Pellicane}
\affiliation{Dipartimento di Scienze Biomediche, Odontoiatriche e delle Immagini Morfologiche e Funzionali, Universit\`a degli Studi di Messina, I-98125 Messina, Italy}
\affiliation{\giuseppe{Discipline of Physics, University of KwaZulu-Natal and National Institute of Theoretical and Computational Sciences (NITheCS), 3209 Pietermaritzburg, South Africa}}
\author{Franz Saija}
\affiliation{Institute for Chemical-Physical Processes, National Research Council (CNR-IPCF), Viale F. Stagno d'Alcontres 37, I-98158 Messina, Italy}
\begin{abstract}
In this work we assess how well a continuous \santi{pseudo-hard-sphere (PHS) potential}, the
Mie(50,49) interaction, can represent a symmetric,
negatively non-additive hard-sphere (NAHS) mixture, using Monte Carlo
equations of state and event-driven molecular dynamics of the true
hard-core mixture as references. At non-additivities $\Delta=-0.1$ and
$-0.2$, the PHS fluid densities exceed the NAHS ones by up to
$0.8\%$, increasing with pressure. Direct solid--fluid coexistence
gives PHS freezing pressures within $0.9\%$ of the hard-sphere value
for the FCC crystal and within $0.25\%$ of the NAHS value for the B1
crystal at $\Delta=-0.3$. As to dynamics, at equal density, PHS fluids diffuse
$15$--$20\%$ faster than the NAHS mixture, and the difference grows on
approaching dynamical arrest. On the other hand, for glassy states the two models have the
same structure beyond contact, but the softer Mie repulsion lowers the
contact values \rev{and the pressure, so that at equal density the NAHS
pressure exceeds the PHS one by $12$--$18\%$,} enlarges the cage,
and lowers the nonergodicity parameters. Varying the repulsive
exponent between 36 and 72 shows that all these differences decrease
monotonically toward the hard-sphere limit, and comparing at equal
pressure instead of equal density removes only a quarter to a third of
the fluid acceleration, but near arrest it removes most of the
difference at the highest pressures studied. As a whole, the PHS model
provides a relatively accurate picture of the structure and
thermodynamics of these hard-sphere mixtures, and a fairly reasonable approximation to
the hard-core dynamics. 
\end{abstract}
\maketitle

\section{Introduction}
Whether a liquid mixture crystallizes or vitrifies on compression or
cooling depends on how its components pack. Amorphous metal alloys
show high strength and elastic limits, and they can be shaped like
polymers.\cite{Schroers2013} They are usually found near deep
eutectics, where Turnbull's argument places the best conditions for
avoiding extended crystals.\cite{Turnbull1969} Binary colloidal
suspensions are used to make either ordered or disordered photonic
materials.\cite{Yadav2025,Hu2021} Most model glass formers use two
ingredients: unequal particle sizes and a preference for unlike
neighbors. The Kob--Andersen Lennard-Jones mixture is the standard
example.\cite{Kob1994} Even so, long simulations show that this
mixture eventually crystallizes, so its glass-forming ability is a
kinetic property.\cite{Pedersen2018} Similar trade-offs between
frustration and crystallization have been mapped for soft-sphere
mixtures as a function of size ratio and composition.\cite{Zhang2015} 

Hard spheres strip this problem down to excluded volume. The
equilibrium state of a hard-sphere system is set by entropy alone. A
one-component hard-sphere fluid crystallizes readily, in colloidal
experiments\cite{Pusey1986} and even from apparently arrested
glasses.\cite{Zaccarelli2009} For this reason, studies of the
hard-sphere glass transition usually rely on size polydispersity or
size asymmetry.\cite{BerthierBiroli2011,Royall2015,Lazaro2019}
Non-additivity offers a
different route. If both species have the same diameter and only the
cross diameter $\sigma_{12}=(1+\Delta)\sigma$ is reduced, the mixture
keeps \giuseppe{the excluded volume between like particles of a one-component fluid}. \santi{Mixing of unlike neighbors is then favored for \giuseppe{excluded-volume} reasons, since \giuseppe{the total excluded volume is reduced and} it allows the mixture to reach higher densities, that is, a smaller Gibbs free energy.} This symmetric negatively
non-additive hard-sphere (NAHS) mixture acts as an entropic analog of
a negative mixing enthalpy. For $\Delta=-0.1$, Punnathanam and Monson
found congruent freezing and a eutectic between substitutionally
disordered FCC and BCC solids, the latter becoming the B2 crystal at
$c=0.5$.\cite{Punnathanam2006} Prestipino \emph{et al.}\cite{PPLS2026}
extended the  
phase map to $\Delta=-0.2$ and $-0.3$. At $\Delta=-0.2$, compression does not
crystallize the fluid over a wide range of intermediate
compositions. At $\Delta=-0.3$ crystallization is recovered at all
compositions, and the equimolar mixture freezes into a B1 (rock-salt)
crystal. This B1 regime sets in when the cross diameter drops below
the size of the octahedral cavity of the FCC lattice,
$(1+\Delta)\le\sqrt2/2$, i.e.\ $\Delta\lesssim-0.293$, which allows
two interpenetrating close-packed FCC lattices. Equal sizes therefore
do not make crystallization unavoidable: a single geometric parameter
can either frustrate crystallization or select a new compound. A
careful choice of non-additivity (geometric constraint) and \giuseppe{relative number of the two species} (compositional constraint) can lead to highly frustrated
states incompatible with crystal phases, as shown in
Ref.~\onlinecite{PPLS2026}. Monte
Carlo approaches, however, carry no physical time, so these results say nothing
about the dynamics of the glass-forming window. Filling this gap is \santi{the} central goal of this article. 

Following the dynamics requires either event-driven molecular dynamics
(EDMD) of the hard-core mixture\cite{Bannerman2011} or a continuous
potential that general-purpose MD codes can integrate. Jover \emph{et
al.}\ proposed the pseudo-hard-sphere potential: a Mie(50,49)
interaction cut at its minimum, shifted, and used at
$k_BT/\epsilon=1.5$.\cite{Jover2012} It reproduces the hard-sphere
fluid and solid equations of state, structure, and transport. The same
work also tested tangent chains and highly size-asymmetric additive
mixtures. With direct coexistence, Espinosa \emph{et al.}\ located the
fluid--FCC transition of this model at a pressure only about 0.6\%
above the hard-sphere value. They thus established the potential as a
tool for studying crystallization.\cite{Espinosa2013} Báez \emph{et
al.}\ derived an essentially equivalent model by matching the
hard-sphere second virial coefficient, which fixes
$k_BT/\epsilon\simeq1.47$ for the same exponent.\cite{Baez2018} That
criterion has since been extended to multicomponent and polydisperse
additive dispersions,\cite{MartinezRivera2023} applied to depletion
forces in colloidal mixtures,\cite{SantosLopez2022} and tested across
fluid--solid coexistence in three and four
dimensions.\cite{BedollaMontiel2025} The closest precedent for the
present problem is a two-dimensional one: a non-additive binary
mixture of pseudo-hard disks used for the inverse design of crystals
and quasicrystals.\cite{BedollaMontiel2024} \santi{It is worth stressing that, at $k_BT/\epsilon=1.5$, the cut-and-shifted Mie(50,49) potential is considerably harder than a pure inverse-power repulsion $r^{-n}$ with a similar exponent, because its large prefactor, $C\simeq135$ in Eq.~(\ref{eq:mie}), makes the repulsion near contact much steeper than that of $r^{-50}$. Measured by the departure of the Boltzmann factor $e^{-\beta u(r)}$ from a hard-sphere step, or by its $10$--$90\%$ width, it is equivalent to an inverse-power potential with $n\approx150$.} 

These validations concern one-component systems, additive mixtures, or
two-dimensional ordering. None of them addresses slow dynamics in a
three-dimensional non-additive mixture, where the mapping faces a
harder test. With $\Delta<0$ the dense packing is controlled by the
short 1--2 contact. Near arrest, small differences in effective
diameter are strongly amplified in relaxation times. Softness itself
is known to shift the balance between crystallization and glass
formation in nearly-hard binary mixtures.\cite{Dasgupta2020} It also
changes how close a soft-sphere glass lies to its hard-sphere
counterpart.\cite{Schmiedeberg2011} A meaningful comparison therefore
needs a hard-core dynamical reference, not just equilibrium data, and
a careful choice of state variables and time units.\cite{Dyre2016} 

\rev{In this work we extend the pseudo-hard-sphere description to symmetric
NAHS mixtures and assess its performance not only in describing the
hard-core thermodynamics and structure (including phase transitions),
but also in the realm of dynamic properties, including arrested
states.} We denote the Mie(50,49) representation of the mixture as \santi{PHS, for pseudo-hard spheres}. \rev{Four aspects deserve to be stressed:} (i)~We
present what to our knowledge is the first
validation of a pseudo-hard-sphere potential for a three-dimensional
non-additive mixture, covering both the glass-forming window
($\Delta=-0.1$ and $-0.2$) and the B1-forming regime ($\Delta=-0.3$)
found in Ref.~\onlinecite{PPLS2026}. The test combines the Monte Carlo
equations of state of that work with direct solid--fluid coexistence
for the FCC and B1 crystals.\cite{Alkemade2026} (ii)~In addition, we
introduce a dynamical benchmark of a pseudo-hard-sphere model against
event-driven simulations of the true hard-core mixture; earlier
validations were limited to equilibrium properties. (iii)~\rev{We
complement the results of the Monte Carlo study of
Ref.~\onlinecite{PPLS2026} with the
first characterization of the dynamics of the NAHS glass-forming
window.} We analyze relaxation, structure, and elastic response of the
glass. (iv)~Finally, \rev{by varying the repulsive exponent, we separate the effects of
softness from the properties of the NAHS mixture itself, and
quantify how far the mapping can be trusted near
arrest.}

The rest of the paper is sketched as
follows. Section~\ref{sec:methods} describes the models and simulation 
methods. Section~\ref{sec:valid} validates thermodynamics, structure,
and coexistence. Section~\ref{sec:glass} analyzes the glass dynamics,
the comparison with EDMD, and the softness
dependence. \rev{Sections~\ref{sec:disc} and~\ref{sec:concl} discuss the
results and present our conclusions.} 

\section{Models and methods}\label{sec:methods}
\subsection{NAHS model and Monte Carlo reference}
We will be dealing with two species that have equal like-particle contact diameters,
$\sigma_{11}=\sigma_{22}=\sigma$, and cross diameter
$\sigma_{12}=(1+\Delta)\sigma$. The composition is $c=N_2/N$. For
$\Delta=-0.1$ and $-0.2$, $N_1=2041$, $N_2=875$, and $c\simeq0.30$;
for $\Delta=-0.3$, $N_1=1750$, $N_2=1166$, and $c\simeq0.40$. All
systems \giuseppe{in one-phase simulations} have $N=2916$. \rev{The results of Ref.~\onlinecite{PPLS2026} are
our primary NAHS reference for thermodynamics and structure, in
particular its $NPT$ equation-of-state data; the two $NVT$ branches
are complementary data from the same work that were not shown
there.} 

\subsection{The PHS Mie(50,49) model}
For a pair of species $i,j$, the cut-and-shifted potential is\cite{Jover2012}
\begin{equation}\label{eq:mie}
u_{ij}(r)=\begin{cases}
C\epsilon\left[(\sigma_{ij}/r)^{50}-(\sigma_{ij}/r)^{49}\right]+\epsilon,&r<r_{\min,ij},\\
0,&r\ge r_{\min,ij},
\end{cases}
\end{equation}
where $r_{\min,ij}=(50/49)\sigma_{ij}$ and
\begin{align}
C&=\frac{n}{n-m}\left(\frac{n}{m}\right)^{m/(n-m)},\\
C\big|_{(n,m)=(50,49)}&=50(50/49)^{49}.
\end{align}
We retain the value of $T^*=k_BT/\epsilon=1.5$ suggested by Jover
\emph{et al.}\cite{Jover2012} for this mapping. The pressure is reported as $\bP=P^*/T^*$, with $P^*=P\sigma^3/\epsilon$. Time is reported in $\tau_{\mathrm{HS}}=\sigma\sqrt{m/(k_BT)}$; hence $t_{\mathrm{HS}}=t^*\sqrt{T^*}$ for $t^*=t\sqrt{\epsilon/m}/\sigma$.

\subsection{MD and MC simulations}
The PHS model was simulated with LAMMPS molecular dynamics, including
$NPT$ compression and fixed-volume trajectories.\cite{Thompson2022}
The NAHS reference uses hard-core Monte Carlo with particle-swap
moves, as in Ref.~\onlinecite{PPLS2026}. \giuseppe{The NAHS structure factors shown here (Fig.~\ref{fig:struct}), which were not reported in Ref.~\onlinecite{PPLS2026}, were computed from our own MC simulations with the MC\_Checkerboard code\cite{MCCheckerboard2026} and analyzed with \texttt{trj\_analysis}.\cite{trjAnalysis2026}} Each tabulated $NPT$ state
comprises typically $10^6$ MC cycles ($5\times10^5$ for equilibration
and $5\times10^5$ for production); statistical errors are comparable
to the symbol size in the figures of Ref.~\onlinecite{PPLS2026}, where
further details can be found. The PHS runs
use a time step $\delta t=0.001\,\tau_{\rm LJ}$ and Nosé--Hoover
thermostat and barostat with relaxation times \santi{$0.1\,\tau_{\rm LJ}$} and $1.0\,\tau_{\rm LJ}$. Each $NPT$ state is equilibrated for
$5\times10^{6}$ steps and sampled for $2\times10^{6}$ steps; mean
densities have block-averaged standard errors of
$\approx5\times10^{-5}$ \giuseppe{(below $0.005\%$ of the mean density)}. Glass states are run at fixed density after
the same equilibration. Dynamical trajectories at fixed density are
run in the microcanonical ensemble after thermostatted equilibration;
runs longer than $10^{6}$ steps use a weak thermostat (relaxation time
$1\,\tau_{\rm LJ}$), because the microcanonical temperature drifts
appreciably over $10^{7}$ steps. Independent glass samples are
obtained by $NPT$ compression of the fluid at rates between $10^{6}$
and $10^{7}$ steps per ramp. 

\subsection{Event-driven Molecular Dynamics of NAHS}
Event-driven molecular dynamics (EDMD) with the DynamO simulation
package\cite{Bannerman2011} supplies an exact 
dynamics reference for the hard-core mixture. We
use $\sigma_{11}=\sigma_{22}=\sigma$, $\sigma_{12}=(1+\Delta)\sigma$,
$m=k_BT=1$, and the hard-sphere time unit $\tau_{\rm
  HS}=\sigma\sqrt{m/k_BT}$; PHS times are converted with $t/\tau_{\rm
  HS}=(t/\tau_{\rm LJ})\sqrt{T^*}$. Each EDMD sample has the same $N$,
composition, and number density as an PHS state. Starting
configurations are either an PHS frame rescaled to the PHS mean
density, with diameters temporarily reduced by the minimum overlap
ratio \giuseppe{of the particle diameters} ($\lambda\simeq0.974$--$0.981$) and then restored by EDMD
compression, or, for the glasses, an overlap-free NAHS MC
configuration. For each glass, three EDMD samples start from the same
MC configuration with independent velocities. The $\Delta=-0.2$
fluid second samples likewise reuse the PHS positions; these samples
measure dynamical noise, not configuration-to-configuration
variation. After thermalization, fluid samples are equilibrated for
$2000\,\tau_{\rm HS}$ and glass samples for $4500$--$5000\,\tau_{\rm
  HS}$, followed by $NVE$ production of $1000\,\tau_{\rm HS}$ (fluids,
frames every $0.5\,\tau_{\rm HS}$) or $200\,\tau_{\rm HS}$ (glasses,
every $0.1\,\tau_{\rm HS}$). The pressure is the collision virial. As
a check, the single-component EDMD pressure at $\phi=0.30$ agrees with
the Carnahan--Starling equation within $0.14\%$. Trajectories are
analyzed with the same code and settings as the PHS trajectories. 

\subsection{Analysis: correlation functions}
For species $\alpha$, the mean-square displacement (MSD) and velocity autocorrelation function (VACF) are
\begin{align}
M_\alpha(t)&=\langle|\mathbf r_i(t)-\mathbf r_i(0)|^2\rangle_{i\in\alpha},\\
Z_\alpha(t)&=\frac{1}{3}\langle\mathbf v_i(t)\cdot\mathbf v_i(0)\rangle_{i\in\alpha}.
\end{align}
The long-time relations $M_\alpha(t)\sim6D_\alpha t$ and $D_\alpha=\int_0^\infty Z_\alpha(t)\,dt$ apply in a diffusive state.\cite{Hansen2013} We use the number-density mode $\rho_{\mathbf Q}(t)=\sum_j e^{i\mathbf Q\cdot\mathbf r_j(t)}$ and define
\begin{align}
F(Q,t)&=N^{-1}\langle\rho_{\mathbf Q}(t)\rho_{-\mathbf Q}(0)\rangle,\\
F_s(Q,t)&=N^{-1}\sum_j\langle e^{i\mathbf Q\cdot[\mathbf r_j(t)-\mathbf r_j(0)]}\rangle.
\end{align}
Thus $F(Q,0)=S_{NN}(Q)$ and $F_s(Q,0)=1$. We define $S(Q,\omega)=(2\pi)^{-1}\int_{-\infty}^{\infty}F(Q,t)e^{i\omega t}\,dt$. Longitudinal and transverse current spectra use the components of $\mathbf j_{\mathbf Q}=\sum_j\mathbf v_j e^{i\mathbf Q\cdot\mathbf r_j}$ parallel and perpendicular to $\mathbf Q$, respectively.\cite{Hansen2013,trjAnalysis2026} \santi{The nonergodicity parameter is the long-time limit of the normalized collective correlator, $f_Q=\lim_{t\to\infty}F(Q,t)/F(Q,0)$, and its self counterpart is $f_Q^{s}=\lim_{t\to\infty}F_s(Q,t)$; nonzero values signal structural arrest. In practice we estimate both from the intermediate-time plateau of the correlators rather than from a single-time ratio.}\cite{Gotze2009} The plateau values and their uncertainties are given in Sec.~\ref{sec:hsdyn} and in the Supplementary Material.

\begin{table}[b]
\caption{Characteristic wavevectors selected from the static-structure analysis, in units of $\sigma^{-1}$.}
\label{tab:q}
\begin{ruledtabular}
\begin{tabular}{lcc}
Feature & $\Delta=-0.1$ & $\Delta=-0.2$\\
Smallest $Q$ & 0.49 & 0.51\\
\santi{Compositional prepeak} & 5.37 & 6.10\\
Principal density peak & 7.82 & 8.13\\
First minimum & 10.75 & 11.20\\
Second-shell peak & 14.17 & 14.20
\end{tabular}
\end{ruledtabular}
\end{table}

\section{Validation of the PHS mapping}\label{sec:valid}
\subsection{Equation of state at $c=0.30$}
\begin{figure*}
\includegraphics[width=\textwidth]{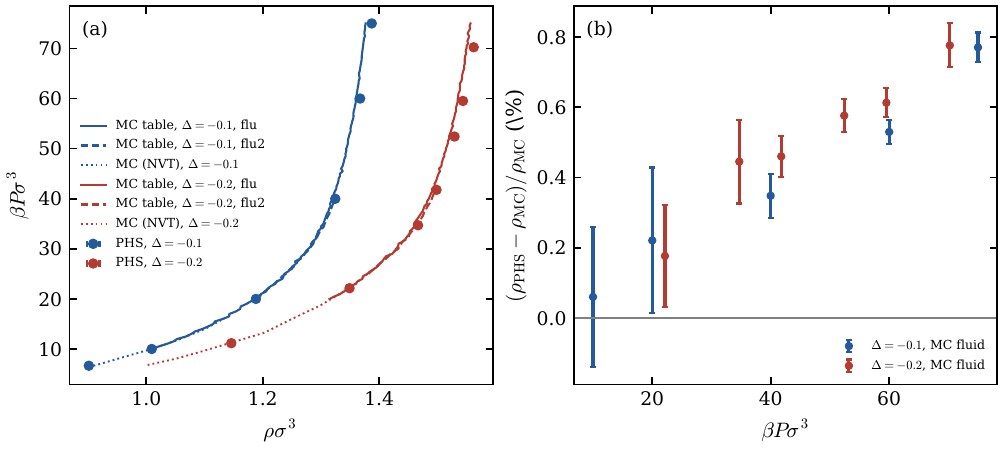}
\caption{Equation of state at $c=0.30$ for $\Delta=-0.1$ and $-0.2$. The deviation panel shows PHS relative to the \giuseppe{fluid-branch data} of Ref.~\onlinecite{PPLS2026}; error bars indicate local fit residual scatter. The $NVT$ branches are complementary data from Ref.~\onlinecite{PPLS2026}.}
\label{fig:eos}
\end{figure*}

Figure~\ref{fig:eos} compares $\bP$ against $\rho\sigma^3$ for
$\Delta=-0.1$ and $-0.2$. Against the fluid \giuseppe{equation-of-state data} of
Ref.~\onlinecite{PPLS2026} \giuseppe{(provided by the coauthors; not tabulated in that paper)}, the PHS density deviations at
$\Delta=-0.1$ and $\bP=10,20,40,60,$ and $75$ are $+0.14$, $+0.06$,
$+0.26$, $+0.59$, and $+0.75\%$ by linear interpolation. At $\Delta=-0.2$ and measured
$\bP=22.15,34.70,41.78,52.41,59.53,$ and $70.24$, they are $+0.20$,
$+0.44$, $+0.51$, $+0.62$, $+0.59$, and $+0.79\%$ by
interpolation. The statistical errors of the Monte Carlo data are 
comparable to the symbol size in the figures of
Ref.~\onlinecite{PPLS2026}, which corresponds to
$\delta(\rho\sigma^3)\approx\pm0.01$ ($\approx0.7\%$); the PHS
deviations are therefore within these errors except at the highest
pressures ($\bP\gtrsim70$). The scatter of neighboring table points
(0.1--0.3\% in density) and the monotonic growth of the deviation with
pressure indicate that the deviation is systematic, as confirmed by
the event-driven hard-sphere pressures (Sec.~\ref{sec:hsdyn}). 

\subsection{Structure}
\begin{figure}
\includegraphics[width=\columnwidth]{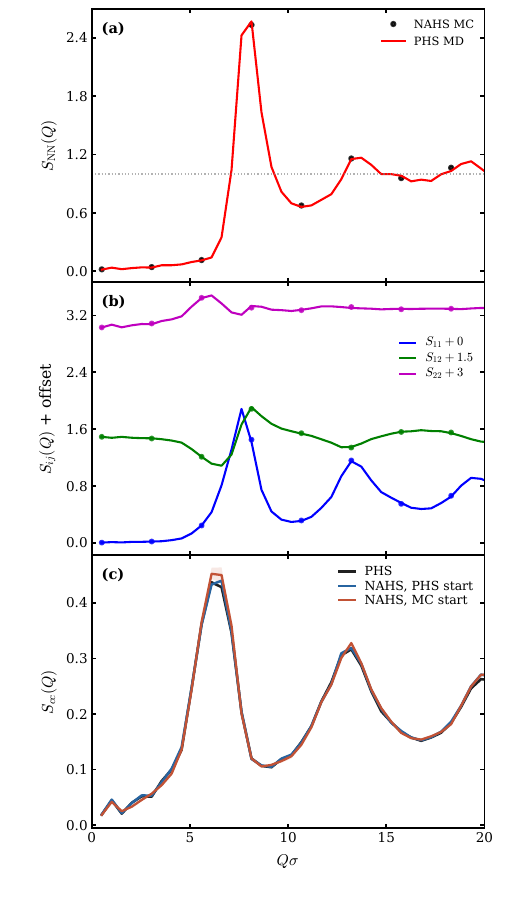}
\caption{Structure of the $\Delta=-0.2$ glass
  ($\rho\sigma^3=1.5442$). (a) Number--number and (b) partial
  structure factors (shifted vertically) for the NAHS mixture from MC
  (points) and for PHS (lines); the PHS glass inherited its initial
  structure from MC. (c) Concentration structure factor for PHS and
  for the NAHS mixture (EDMD) started from the PHS and from the MC
  configuration. The partial pair distributions near contact are shown
  in Fig.~\ref{S-fig:grcontact02} of the supplementary material.} 
\label{fig:struct}
\end{figure}

Figure~\ref{fig:struct} compares the structure factors of the two
models in the $\Delta=-0.2$ glass; the partial pair distributions near
contact are shown in Fig.~\ref{S-fig:grcontact02} of the supplementary
material; the \giuseppe{structural quantities for the} $\Delta=-0.1$ glass, which \giuseppe{behave} in the same way, \giuseppe{are}
shown in Figs.~\ref{S-fig:struct01} and~\ref{S-fig:gr01} of the
supplementary material. The glass-state PHS runs were started from
NAHS MC configurations. Agreement in these states therefore \giuseppe{might} partly
\giuseppe{reflect} the shared initial configuration. Two tests remove this
ambiguity (Sec.~\ref{sec:hsdyn}). In the ergodic fluids at
$\bP\approx20$--$22$, where the EDMD samples have diffused far from
their initial configurations, the PHS and NAHS structures differ
slightly but significantly: the main peak of $S_{NN}(Q)$ is
$0.5$--$1.3\%$ lower in PHS and the contact peaks of $g_{ij}(r)$ are
$10$--$15\%$ lower, as expected from
the softness of the PHS potential. In contrast, \santi{the Bhatia--Thornton concentration--concentration structure factor, $S_{cc}(Q)=x_2^2S_{11}(Q)+x_1^2S_{22}(Q)-2x_1x_2S_{12}(Q)$, with $x_i=N_i/N$ and $S_{ij}(Q)=N^{-1}\langle\rho^{(i)}_{\mathbf Q}\rho^{(j)}_{-\mathbf Q}\rangle$ (so that $S_{NN}=S_{11}+S_{22}+2S_{12}$ and $S_{cc}\to x_1x_2$ at large $Q$), which describes the compositional order,} agrees within the statistical
noise. In the glasses, four independently compressed PHS samples per
$\Delta$ and their EDMD counterparts give the sample-to-sample
scatter, within which the MC-started glasses lie. This confirms the
overall agreement between the PHS MD and the corresponding MC results
for the NAHS.  

\subsection{Liquid--solid transition at $\Delta=-0.3$, $c=0.40$}
\begin{figure}
\includegraphics[width=\columnwidth]{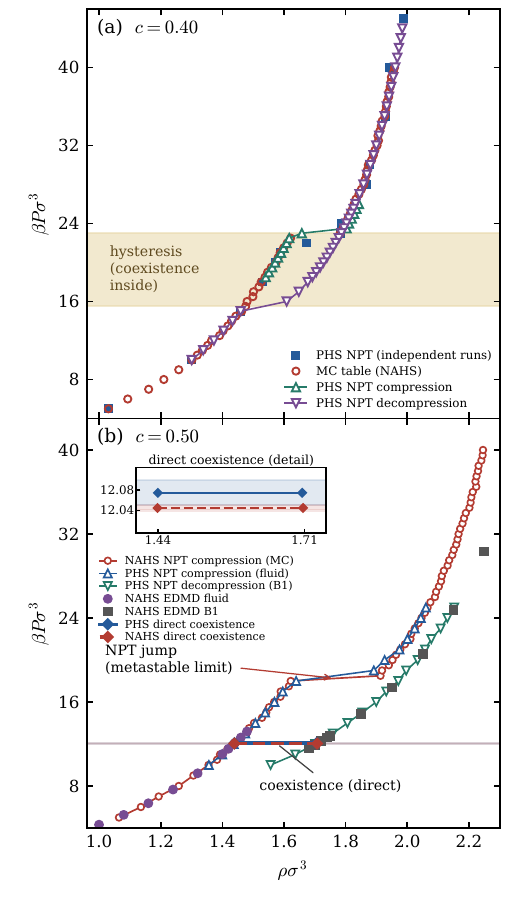}
\caption{Equation of state of the $\Delta=-0.3$ mixture near
  freezing. (a) $c=0.40$: independent PHS $NPT$ runs, sequential PHS
  compression and decompression branches, and the $NPT$ data of
  Ref.~\onlinecite{PPLS2026} (tabulated coauthor data); the shaded
  band is the hysteresis loop within which coexistence lies. (b)
  $c=0.50$: $NPT$ compression branch of Ref.~\onlinecite{PPLS2026},
  sequential PHS fluid and B1 branches, and EDMD bulk points for the
  NAHS mixture. Horizontal tie lines mark solid--fluid coexistence
  from direct-coexistence simulations (Sec.~\ref{sec:coex}), far below
  the $NPT$ density jump. Beyond the jump, the compressed states are
  only partially crystalline; the B1 branches started from the lattice
  represent the stable solid above coexistence.} 
\label{fig:ls}
\end{figure}

In independent PHS $NPT$ runs at $c=0.40$, the density jumps between
$\bP\simeq21$ and 23 into a B1-like solid branch
[Fig.~\ref{fig:ls}(a)]. In the figure a reference curve from
Ref.~\onlinecite{PPLS2026} is  included for comparison. This
compression path alone 
does not establish coexistence pressures or a hysteresis width. The
coexistence test of Sec.~\ref{sec:coex} shows that such jumps occur
deep in the metastable region. 

\subsection{Direct solid--fluid coexistence}\label{sec:coex}
Compression and decompression paths are kinetic, so we test the
mapping directly on solid--fluid coexistence, following Espinosa
\emph{et al.}\cite{Espinosa2013} \rev{and using direct solid--fluid
coexistence simulations. As a testing ground, we first consider the}
one-component FCC 
crystal, for which reference values exist, and the equimolar B1 (NaCl)
crystal at $\Delta=-0.3$, $c=0.5$, where the crystal and the fluid can
share the same composition. Each system has $N=5184$ particles: an
equilibrated crystal slab with 2592 particles in contact, through two
periodic interfaces along $z$, with a fluid slab of 2592
particles. The fluid is prepared separately: it is expanded along $z$
to $\rho\sigma^3=0.80$, melted, and recompressed at the target
pressure with the lateral box lengths of the crystal, and it is
accepted only if its crystalline fraction is below $5\%$. Bond-orientational order is measured with the Steinhardt parameters\cite{Steinhardt1983}\giuseppe{:} \santi{for each particle $i$, $q_{lm}(i)=N_b(i)^{-1}\sum_{j}Y_{lm}(\hat{\mathbf r}_{ij})$ is computed from its $N_b(i)$ neighbors within a cutoff $r_c$, and $q_l(i)=[4\pi/(2l+1)\sum_{m=-l}^{l}|q_{lm}(i)|^2]^{1/2}$ is its rotationally invariant combination;} and the reported $Q_l$ is the average of $q_l$ over particles and configurations; no neighbor averaging of the $q_{lm}$ is applied. For the B1 solid we take $r_c=0.84\sigma$, which lies between the unlike first shell and the like second shell, so that only the octahedral unlike contacts are counted; the resulting values are discussed in the supplementary material. \giuseppe{For the solid--fluid coexistence analysis we use separately calibrated bond-correlation criteria, described next.}
Crystalline
particles are identified with a bond-orientational criterion
calibrated on equilibrated fluids and crystals: connected $q_6$
bonds\cite{tenWolde1996} for FCC, \giuseppe{whereas}, because the expanded B1 lattice
defeats that criterion, a first-shell $q_4$ bond count averaged over
neighbors \giuseppe{is used} for B1. Each criterion assigns at most $1\%$ of particles in
equilibrated fluids and more than $96\%$ in thermal crystals to the
solid \giuseppe{(this calibration criterion is distinct from the $5\%$ limit above, which only serves to accept the initial fluid slab)}. 

\giuseppe{The two slabs are joined by stacking them along $z$, with a gap of $1.05\sigma$ (PHS) or $1.15\sigma$ (NAHS) at each of the two interfaces, after verifying that there are no overlaps. For NAHS, the fluid slab is prepared by EDMD compression to the target density, with its lateral box lengths matched to those of the crystal, and the same acceptance criterion on its crystalline fraction applies to PHS and NAHS fluids alike; the joined NAHS system is then equilibrated for $200\,\tau_{\rm HS}$ before production.} For PHS, each joined system is run at fixed $\bP$ with $NPT$ dynamics along $z$ only, and is classified as frozen or melted when its crystalline fraction exceeds $0.8$ or falls below $0.2$; the coexistence pressure lies where the outcome changes. For the NAHS mixture, EDMD at fixed volume brings the system to coexistence, and we take the time-averaged pressure of eight independent samples whose crystal slab, at least $3\sigma$ thick, survives the whole run. The results are collected in Table~\ref{tab:coex}.

\begin{table}
\caption{Solid--fluid coexistence pressures $\bP_c$ from direct coexistence ($N=5184$). For PHS, the bracket between the highest pressure at which all runs melted and the lowest at which all runs froze; for NAHS (EDMD), the mean over eight samples and its standard error. FCC: $\Delta=0$; B1: $\Delta=-0.3$, $c=0.5$. The last column is the relative difference.}
\label{tab:coex}
\begin{ruledtabular}
\begin{tabular}{lccc}
Crystal & PHS & NAHS (EDMD) & Diff.\\
\hline
FCC & $11.65\pm\rev{0.05}$ & $11.547\pm0.010$ & $+0.9\%$\\
B1 & $12.075\pm0.025$ & $12.045\pm0.005$ & $+0.25\%$\\
\end{tabular}
\end{ruledtabular}
\end{table}

For FCC, the PHS runs melt in all cases at $\bP\le11.60$ and freeze
in all cases at $\bP\ge11.70$, with mixed outcomes at
$11.625$--$11.675$ (54 runs, up to $7000\,\tau$), \rev{so that the coexistence pressure is bracketed within $11.65\pm0.05$}. This reproduces the
value $11.65(1)$ reported for this model by Espinosa \emph{et
al.}\cite{Espinosa2013} The EDMD hard-sphere value, $11.547\pm0.010$,
agrees with the hard-sphere reference values quoted there; a second
EDMD set with a thinner crystal slab gives $11.557\pm0.016$. For B1,
all 10 PHS runs at $\bP\le12.05$ melt and all 56 runs at
$\bP\ge12.10$ freeze, except one run at $12.10$ that remained
inconclusive; the transition is thus located within $\pm0.025$. The PHS
coexistence pressure exceeds the NAHS value by $0.9\%$ for FCC and by
only $0.25\%$, within the uncertainty, for B1. The mapping that
reproduces the fluid equation of state at $T^*=1.5$ therefore also
reproduces the freezing transition of the non-additive mixture. The
coexisting densities follow from the bulk equations of state of each
phase at the measured coexistence pressure. For the B1 crystal at
$c=0.5$ they are $\rho_f\sigma^3=1.440$ and $\rho_s\sigma^3=1.706$ for
PHS, and $1.439$ and $1.707$ for the NAHS mixture, a density change
of about $19\%$ on freezing. In the EDMD slabs, the interior of the
fluid region has $\rho\sigma^3=1.438\pm0.004$, in agreement with the
bulk value; the density of the crystal slab cannot be measured
independently in these runs, because its lateral dimensions are fixed
by the initial lattice. 

The bulk paths are far from coexistence. In the $c=0.5$ PHS mixture,
the fluid compressed in steps of $\Delta\bP=1$ (50\,$\tau_{\rm LJ}$
per step) first crystallizes between $\bP=18$ and 19, as does the
$NPT$ branch of Ref.~\onlinecite{PPLS2026} between $18$ and $18.5$,
whereas the B1 crystal survives decompression down to $\bP=10$. A
fluid prepared at $\bP=18$ crystallized within $150\,\tau_{\rm
  LJ}$. As Fig.~\ref{fig:ls}(b) shows, the $NPT$ density jump
therefore lies about $50\%$ above the coexistence pressure,
$\bP_c\approx12.1$, and marks the limit of metastability of the
compressed fluid rather than the freezing transition. The states
reached beyond the jump are also not the equilibrium solid. We
classify a particle as B1-crystalline if it has six unlike nearest
neighbors in octahedral geometry ($q_4>0.6$) and no like nearest
neighbor. By this criterion, the PHS samples compressed from the
fluid are only $60$--$77\%$ crystalline at $19\le\bP\le25$ and
$4$--$6\%$ less dense than samples started from the B1 lattice at the
same pressure, which are $83$--$95\%$ crystalline; neither contains
like nearest-neighbor pairs, so the deficit reflects non-crystalline
regions and defects rather than \santi{compositional disorder}. The compression
branch of Ref.~\onlinecite{PPLS2026} lies on the same low-density side
of the EDMD B1 points (by $3$--$4\%$ at $\bP\approx30$). Above the
coexistence pressure the B1 crystal has the lower Gibbs free energy:
since $d\mu=v\,dP$ along an isotherm, $\beta\mu_{\rm
  fluid}-\beta\mu_{\rm B1}=\int_{P_c}^{P}(v_{\rm fluid}-v_{\rm
  B1})\,d(\beta P')$ grows from zero at $P_c$, and with $v_{\rm
  fluid}-v_{\rm B1}\approx0.11\sigma^3$ it reaches $\approx0.5\,k_BT$
per particle at $\bP\approx17$. The partially crystalline states
beyond the $NPT$ jump are therefore metastable with respect to the
branch started from the B1 lattice, which represents the stable solid
above $P_c$ (below $P_c$ it is a superheated solid that survives on
the simulation time scale). At $c=0.40$, the composition of
Fig.~\ref{fig:ls}, sequential compression in steps of $\Delta\bP=0.5$
($100\,\tau_{\rm LJ}$ per step) keeps the fluid up to $\bP=22.5$ and
yields a B1-like solid from $\bP=23.5$ on, in agreement with the $NPT$
branch of Ref.~\onlinecite{PPLS2026}, which jumps between $22.5$ and
$23$. On decompression from $\bP=45$, the solid survives down to
$\bP=16$ and melts between $16$ and $15$. The coexistence pressure at
$c=0.40$ thus lies within the hysteresis loop
$15.5\lesssim\bP\lesssim23$, shaded in Fig.~\ref{fig:ls}(a). Direct
coexistence at this composition would require a non-stoichiometric
crystal; thus, the equimolar B1 crystal at $c=0.5$ is a cleaner test of
the mapping.

\section{Glass dynamics}\label{sec:glass}
Here we extend the results of Ref.~\onlinecite{PPLS2026}, \rev{where crystal
suppression was reported near intermediate compositions for
moderate negative non-additivity. To that aim we resort to EDMD
and validate the use of the PHS model to describe the dynamics of the
NAHS mixture.}

\subsection{Single-particle motion}
\begin{figure}
\includegraphics[width=\columnwidth]{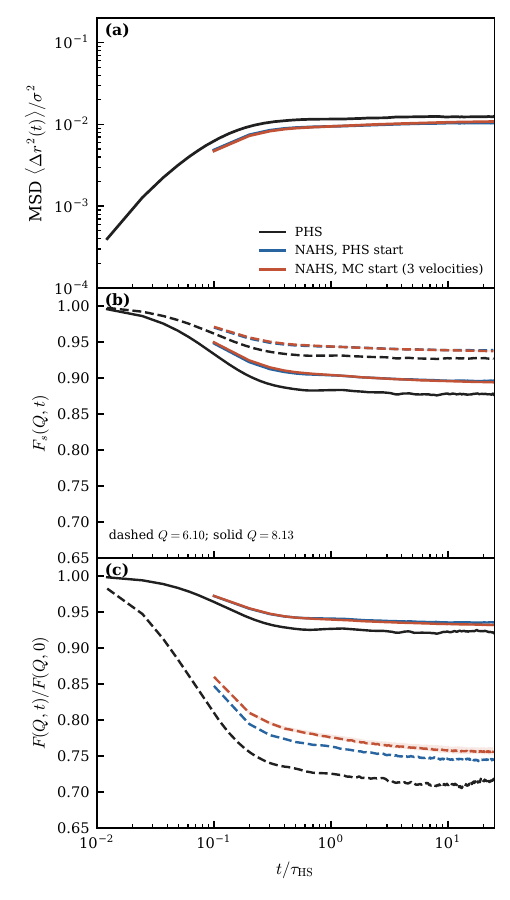}
\caption{Dynamics of the $\Delta=-0.2$ glass versus time in hard-sphere units. (a) Mean-square displacement; (b) self and (c) normalized collective intermediate scattering functions at the \santi{compositional-order} wave number $Q\sigma=6.10$ (dashed) and at the main peak, $Q\sigma=8.13$ (solid). PHS is compared with the NAHS mixture (EDMD) started from the PHS configuration and from the MC configuration (three velocity realizations; mean and range).}
\label{fig:dynamics}
\end{figure}

Figure~\ref{fig:dynamics}(a) shows the mean-square displacement in the
$\Delta=-0.2$ glass ($\rho\sigma^3=1.5442$) for PHS and for the NAHS
mixture started from the PHS and from the MC configuration; the
$\Delta=-0.1$ glass ($\rho\sigma^3=1.3707$) behaves in the same way
(Fig.~\ref{S-fig:dyn01} of the supplementary material). Within the
analysis window ($24.5\,\tau_{\rm HS}$) all samples are arrested: the
MSD reaches a plateau of order $10^{-2}\sigma^2$ after
$t\approx0.5\,\tau_{\rm HS}$ and grows only slowly afterwards. The
PHS plateau is $26$--$36\%$ higher at $\Delta=-0.1$ and $15$--$20\%$
higher at $\Delta=-0.2$ (Sec.~\ref{sec:hsdyn}). Long-time diffusion is
not resolved in the glasses. The slope of the MSD over the second half
of the window gives upper bounds $D\lesssim4\times10^{-6}$ (PHS) and
$D\lesssim1.4\times10^{-6}\,\sigma^2/\tau_{\rm HS}$ (NAHS) at
$\Delta=-0.1$, and $D\lesssim1.3\times10^{-6}$ and
$2\times10^{-7}\,\sigma^2/\tau_{\rm HS}$ at $\Delta=-0.2$; the long
PHS run at $\Delta=-0.1$, $\bP\approx40$ gives
$D\lesssim2\times10^{-7}\,\sigma^2/\tau_{\rm HS}$ over
$4900\,\tau_{\rm HS}$. These bounds lie at least three orders of
magnitude below the fluid values at $\bP\approx20$
($D\approx5\times10^{-3}\,\sigma^2/\tau_{\rm HS}$). 

\subsection{Intermediate scattering functions, $F(Q,t)$, $F_s(Q,t)$,
  and nonergodicity} 
The $Q$ choices in Table~\ref{tab:q} sample long-wavelength,
\santi{compositional-order}, and cage-scale
relaxation. Figures~\ref{fig:dynamics}(b) and~(c) show the self and
collective intermediate scattering functions, $F_s(Q,t)$ and
$F(Q,t)/F(Q,0)$ at the \santi{compositional-order} and main-peak wave numbers. As
consistency checks of the analysis, $F_s(Q,0)=1$ within $10^{-3}$ at
every $Q$, and $F(Q,0)/S_{NN}(Q)$ lies between $0.99$ and $1.05$, the
residual reflecting the interpolation of $S_{NN}$ on the discrete $Q$
grid. All correlators reach a plateau within $1\,\tau_{\rm HS}$ and do
not decay within the window. We take the nonergodicity parameters
$f_Q$ as plateau averages over $2\le t/\tau_{\rm HS}\le20$, with
uncertainties from block averages. At the main peak,
$f_Q^{s}=0.910\pm0.001$ (PHS) against $0.928$--$0.933$ (NAHS) at
$\Delta=-0.1$, and $0.878\pm0.001$ against $0.896$--$0.897$ at
$\Delta=-0.2$; the collective $F(Q,t)$ shows the same ordering at
every $Q$. Along the fluid branch the PHS main-peak relaxation time
grows from $\tau_\alpha=1.7$ ($\Delta=-0.1$, $\bP=20$) and
$1.6\,\tau_{\rm HS}$ ($\Delta=-0.2$, $\bP=22$) to
$2.9\times10^{3}\,\tau_{\rm HS}$ at $\Delta=-0.2$, $\bP\approx35$, and
exceeds $5\times10^{3}\,\tau_{\rm HS}$ at $\bP\approx40$--$42$
(Sec.~\ref{sec:hsdyn}). 

\subsection{Comparison with true hard-sphere dynamics}\label{sec:hsdyn}
\paragraph*{Pressure at matched density.}
Table~\ref{tab:hsp} compares the EDMD pressure of the NAHS mixture
with the PHS pressure at the same number density. The EDMD pressures
agree with the tables of Ref.~\onlinecite{PPLS2026} within
$0.1$--$0.9\%$ for $\bP\le42$, except at $\Delta=-0.2$, $\bP\simeq35$,
where the table is locally noisy and both systems are close to arrest
($-2\%$). \rev{At equal density the NAHS pressure exceeds the PHS
  pressure (the $NPT$ values of Table~\ref{tab:hsp}) by about $1\%$ at
  $\bP\simeq20$, $2\%$ at $\bP\simeq35$, $4$--$5\%$ at $\bP\simeq40$,
  and $12$--$18\%$ in the glasses. Throughout this paper, pressure
  differences between the models are given in this way, as $(P_{\rm
    NAHS}-P_{\rm PHS})/P_{\rm PHS}$; relative to $P_{\rm NAHS}$, the
  glass values would be $11$--$15\%$.} Because the equation of state
is steep, these pressure differences correspond to the small density
deviations of Fig.~\ref{fig:eos}: at $\bP=40$ ($\Delta=-0.1$) the
local slope $d\ln P/d\ln\rho\approx11$ converts a $+0.4\%$ density
deviation, relative to the smoothed table of
Ref.~\onlinecite{PPLS2026}, into a $\approx4\%$ pressure deviation. In
the glasses, where both systems are arrested, the pressure is
controlled by the contact values of $g_{ij}(r)$, which the soft Mie
repulsion lowers (see below).  
\begin{table}
\caption{Pressure of the NAHS mixture from EDMD, $\beta P_{\rm
    HS}\sigma^3$, at the mean density of each PHS state, compared
  with the PHS pressure and with the tables of
  Ref.~\onlinecite{PPLS2026}. For the $\Delta=-0.2$ fluids two EDMD
  samples are listed; for the glasses, the sample started from the
  PHS configuration and the range of three samples started from a
  single MC configuration with different velocities.} 
\label{tab:hsp}
\begin{ruledtabular}
\begin{tabular}{lcccc}
$\Delta$ & $\rho\sigma^3$ & PHS & EDMD & MC, Ref.~\onlinecite{PPLS2026}\\
\hline
$-0.1$ & 1.1883 & 20.00 & 20.19 & 20.21\\
$-0.1$ & 1.3246 & 39.99 & 42.05 & 41.69\\
$-0.1$ & 1.3707 & 64.5 & 75.9; 76.1--76.2 & --\\
$-0.2$ & 1.3492 & 22.15 & 22.33, 22.36 & 22.35\\
$-0.2$ & 1.4669 & 34.70 & 35.40, 35.29 & 36.10\\
$-0.2$ & 1.4985 & 41.78 & 43.57, 43.63 & 43.84\\
$-0.2$ & 1.5442 & 60.3 & 67.8; 70.0--70.4 & --\\
\end{tabular}
\end{ruledtabular}
\end{table}

\paragraph*{Fluid dynamics.}
In the two fluid states that relax completely within the analysis
window ($\Delta=-0.1$, $\bP=20$; $\Delta=-0.2$, $\bP=22$), we compare
the EDMD trajectories with PHS trajectories run at the same density
in the microcanonical ensemble and stored every $0.049\,\tau_{\rm
  HS}$. The PHS pressures in these runs, $\bP=19.98$ and $22.14$,
reproduce the $NPT$ values. The PHS mixture relaxes faster, and by a
factor that hardly depends on wave number. The ratio of structural
relaxation times $\tau_\alpha(Q)$, defined by
$F_s(Q,\tau_\alpha)=e^{-1}$, is $\tau_\alpha^{\rm HS}/\tau_\alpha^{\rm
  PHS}=1.16$, $1.16$, and $1.22$ at $Q\sigma=5.37$, $7.82$, and
$10.75$ for $\Delta=-0.1$, and $1.23$, $1.22$, and $1.21$ at
$Q\sigma=6.10$, $8.13$, and $11.20$ for $\Delta=-0.2$. The diffusion
coefficients obtained over the same time window give the same factors,
$D_{\rm PHS}/D_{\rm HS}=1.16$ and $1.21$. At the main peak,
$\tau_\alpha=1.73$ (PHS) and $2.01\,\tau_{\rm HS}$ (NAHS) for
$\Delta=-0.1$, and $1.58$ and $1.93\,\tau_{\rm HS}$ for
$\Delta=-0.2$. In these fluid states, therefore, the PHS dynamics is
the NAHS dynamics with a uniformly accelerated clock, consistent with
the slightly lower PHS pressure at equal density. Comparing instead
at equal pressure, with EDMD runs at the slightly lower NAHS densities
$\rho\sigma^3=1.18598$ and $1.34670$ that reproduce the PHS pressures
($\bP=19.99$ and $22.15$--$22.17$), reduces the factors to
$1.12$--$1.13$ ($\tau_\alpha$) and $1.12$ ($D$) at $\Delta=-0.1$, and
to $1.18$ and $1.15$ at $\Delta=-0.2$, still independent of $Q$. About
a quarter to a third of the acceleration thus reflects the state-point
matching, and the remainder is intrinsic to the softer potential. 

\paragraph*{Dense states and glasses.}
At $\Delta=-0.1$, $\bP=40$ and at $\Delta=-0.2$, $\bP=35$ and $42$ we
extended both the PHS runs (weakly thermostatted, $20\,000\,\tau_{\rm
  LJ}$) and the EDMD runs ($10\,000\,\tau_{\rm HS}$), and analyzed
them over windows of $\approx5000\,\tau_{\rm HS}$. Two robust
differences appear. First, the PHS cage is somewhat softer: the MSD on the
plateau, at $t=10\,\tau_{\rm HS}$, is $0.0197$, $0.0343$, and
$0.0231\sigma^2$ in PHS against $0.0171$, $0.031$, and
$0.021\sigma^2$ in the NAHS mixture, and the plateau of $F_s(Q,t)$ is
correspondingly lower. Second, the PHS mixture leaves the plateau
earlier. Only at $\Delta=-0.2$, $\bP=35$ does $F_s$ decay through
$e^{-1}$ within the window at large wave number: at $Q\sigma=11.20$,
$\tau_\alpha=805\,\tau_{\rm HS}$ for PHS against $1790$ and
$2046\,\tau_{\rm HS}$ for two NAHS runs, a ratio of $2.2$--$2.5$; at
the main peak, $\tau_\alpha=2862\,\tau_{\rm HS}$ for PHS, while the
NAHS correlator is still at $0.41$--$0.50$ after $4800\,\tau_{\rm
  HS}$. At $\bP=42$ ($\Delta=-0.2$) and $\bP=40$ ($\Delta=-0.1$)
neither model relaxes, but PHS decays further (to $0.667$ against
$0.692$--$0.766$, and to $0.774$ against $0.831$) and its MSD at the
end of the window is $1.2$--$1.9$ and $1.7$ times larger. The speed-up
of PHS relative to NAHS thus grows from $1.16$--$1.22$ in the fluid
to about $2$ near arrest. Near arrest single trajectories differ
considerably (the two NAHS runs at $\bP=42$ end with MSDs of $0.034$
and $0.055\sigma^2$); since the second NAHS runs share their starting
positions with the first, this scatter is dynamical, and it sets the
uncertainty of the factor of $2$. 

\paragraph*{Dense states at equal pressure.}
\rev{At equal density the NAHS pressure exceeds the pressure of the fixed-density (microcanonical) PHS runs, $\bP=39.79$, $34.06$, and $41.33$, which lie slightly below the $NPT$ values of Table~\ref{tab:hsp}, by $5$--$6\%$ at $\bP\approx40$--$42$ and by $3$--$4\%$ at $\bP\approx35$.} We therefore repeated the dense EDMD runs at the slightly lower densities $\rho\sigma^3=1.31879$ ($\Delta=-0.1$), $1.46409$, and $1.49242$ ($\Delta=-0.2$), $0.2$--$0.4\%$ below the PHS densities, which give $\bP_{\rm HS}=39.91$, $34.14$--$34.52$, and $41.47$--$41.56$ against $39.79$, $34.06$, and $41.33$ for the PHS runs. The trajectories cover $10\,000\,\tau_{\rm HS}$ after $2000\,\tau_{\rm HS}$ of equilibration and are analyzed over the same windows. At $\bP\approx40$ and $42$ this small density shift brings the NAHS dynamics within the trajectory scatter of PHS: at the end of the window, $F_s$ at the main peak is $0.776$ against $0.774$ for PHS ($\Delta=-0.1$) and $0.686$--$0.721$ against $0.667$ ($\Delta=-0.2$), and the MSD is $0.031$ against $0.032\sigma^2$ and $0.044$--$0.055$ against $0.063\sigma^2$; the plateau MSDs also approach the PHS values. At $\bP\approx35$ ($\Delta=-0.2$) the PHS mixture remains clearly faster: $\tau_\alpha(Q\sigma=11.20)=2078$ and $4476\,\tau_{\rm HS}$ for the two NAHS runs, against $805\,\tau_{\rm HS}$, and final MSDs of $0.15$--$0.29\sigma^2$ against $0.42\sigma^2$. The two NAHS runs at this state differ by a factor of two in $\tau_\alpha$. The slower one starts from the end of the earlier extension run and is therefore older, which shows that the hard-core mixture still ages on the $10^4\,\tau_{\rm HS}$ scale. Near arrest, the steep density dependence of the dynamics thus makes the equal-density acceleration sensitive to the small pressure difference between the models: at $\bP\approx40$--$42$ most of the apparent speed-up disappears at equal pressure, whereas at $\bP\approx35$ a factor of $2$ or more remains, with an uncertainty set by aging and trajectory scatter.

In the $\Delta=-0.2$ glass ($\rho\sigma^3=1.5442$), the EDMD samples remain arrested over $4700\,\tau_{\rm HS}$ (MSD $\le0.012\sigma^2$). Beyond the first peak, $S_{NN}(Q)$ and the partial $g_{ij}(r)$ of PHS and NAHS coincide. The first-peak heights differ: the hard-core contact peaks reach $g_{11}\approx12$ and $g_{12}\approx12$, whereas the Mie repulsion lowers them to about $9$--$10$; this accounts for the lower PHS pressure. To separate model effects from configuration-to-configuration variation, we prepared four further PHS glasses at each $\Delta$ by compressing the PHS fluid at different rates, and converted each to an EDMD hard-core sample. All are amorphous (crystalline fraction below $0.5\%$). The structure of the MC-started glasses lies within the spread of the independent ones: for example, the $S_{cc}$ peak spans $0.431$--$0.456$ (PHS) and $0.428$--$0.459$ (NAHS) for the independent glasses at $\Delta=-0.2$, against $0.437$--$0.464$ for the MC-started ones, and the contact peaks agree to within $3\%$. The independent glasses have lower pressures, which decrease with decreasing compression rate (from $59.9$ to $58.6$ for PHS and from $66.8$ to $65.1$ for NAHS at $\Delta=-0.2$), whereas the MC configuration gives a NAHS pressure about $5\%$ higher ($70.0$--$70.4$). \rev{For the independently compressed glasses the ratio of NAHS to PHS pressure is the same for every sample, $1.15$ at $\Delta=-0.1$ and $1.11$ at $\Delta=-0.2$; for the MC-started configuration, which has the higher NAHS pressure, it is $1.18$ and $1.16$--$1.17$ (Table~\ref{tab:hsp}).} In the arrested state the PHS cage is again softer, with a plateau MSD of $0.0124\sigma^2$ against $0.0104$--$0.0109\sigma^2$, and nonergodicity parameters $f_Q=0.927$, $0.878$, and $0.788$ at $Q\sigma=6.10$, $8.13$, and $11.20$, against $0.937$--$0.939$, $0.894$--$0.897$, and $0.814$--$0.820$ for the NAHS glasses.

The $\Delta=-0.1$ glass ($\rho\sigma^3=1.3707$) shows the same picture
with a stronger softness effect. The EDMD samples remain arrested over
$5200\,\tau_{\rm HS}$ (MSD $\le0.004\sigma^2$ during
equilibration). The hard-core contact peaks reach $g_{11}\approx12.4$
and $g_{12}\approx15.4$, against $9.1$ and $11.6$ in PHS,
\rev{consistent with the NAHS pressure being $17.7\%$ higher than the
  PHS one.} Here the two independent starting configurations have the
same \santi{compositional order}: the $S_{cc}(Q)$ curves coincide, with peak values
of $0.396$--$0.400$, and their pressures differ by only
$0.3$--$0.4\%$. The main peak of $S_{NN}(Q)$ is $2$--$3\%$ lower for
the MC configuration ($3.50$--$3.53$ against $3.60$--$3.61$); this is
again a comparison of single configurations. The PHS cage is softer
than at $\Delta=-0.2$: the plateau MSD at $t=20\,\tau_{\rm HS}$ is
$0.0098\sigma^2$ against $0.0072$--$0.0078\sigma^2$ ($26$--$36\%$
larger), and $f_Q=0.955$, $0.909$, and $0.838$ at $Q\sigma=5.37$,
$7.82$, and $10.75$, against $0.964$--$0.967$, $0.926$--$0.932$, and
$0.867$--$0.877$. In both glasses, therefore, the PHS and NAHS
structures agree beyond contact, while the softer Mie repulsion lowers
the contact values, the pressure, and the nonergodicity parameters,
and enlarges the cage. The structural comparison is shown in
Fig.~\ref{fig:struct}(c) and Fig.~\ref{S-fig:grcontact02} of the
supplementary material for $\Delta=-0.2$, and in
Figs.~\ref{S-fig:struct01} and~\ref{S-fig:grcontact01} for $\Delta=-0.1$.

\santi{In summary, the comparison with the hard-core mixture is quantitative for the structure beyond contact and for the compositional order, in the fluid and in the glass alike. The dynamics is reproduced up to a time-scale factor: in the fluid a single, wave-number-independent factor of $1.16$--$1.22$ ($1.12$--$1.18$ at equal pressure) maps one model onto the other, whereas near arrest the factor grows to about two at equal density and becomes sensitive to the small pressure difference between the models. In the glasses, the softer repulsion lowers the contact values and the pressure by $12$--$18\%$, enlarges the cage by $14$--$36\%$, and lowers the nonergodicity parameters by $1$--$4\%$. These differences are systematic rather than random, and Sec.~\ref{sec:soft} shows that they are softness effects that vanish smoothly in the hard-sphere limit.} 

\subsection{Dependence on non-additivity}\label{sec:nonadd}
\rnew{The results above refer to two non-additivities, and the comparison between $\Delta=-0.1$ and $-0.2$ is scattered over several subsections and the supplementary material. We collect it here for the PHS model, for which we have the equation of state, the cage dynamics, and the elastic response of both glasses (Table~\ref{tab:nonadd}). Only two values of $\Delta$ are available, so what follows describes a trend between two systems rather than a functional dependence, and each glass is a single state point.

\emph{Packing and compressibility.} At equal pressure the more negative non-additivity gives a denser fluid: from the PHS equations of state, the density at $\bP=20$, 40, and 60 is $11$--$13\%$ higher at $\Delta=-0.2$ than at $\Delta=-0.1$. The dimensionless isothermal compressibility $\rho k_BT\kappa_T$, obtained from the local slope of the equation of state, is also larger at $\Delta=-0.2$, by about $20$--$50\%$ depending on pressure; the slope comes from only a handful of state points, and we estimate its uncertainty at $10$--$15\%$. The Monte Carlo equations of state of the hard-core mixture of Ref.~\onlinecite{PPLS2026} give the same ordering, with values within $20\%$ of the PHS ones (Table~\ref{tab:nonadd}), as do the independent fits used in the elastic analysis of the glasses ($\rho k_BT\kappa_T=1.4\times10^{-3}$ and $1.9\times10^{-3}$ at $\Delta=-0.1$ and $-0.2$). Equivalently, the isothermal sound speed, $c_{\rm iso}=(k_BT/m)^{1/2}(\rho k_BT\kappa_T)^{-1/2}$ in terms of this quantity, is higher at $\Delta=-0.1$ in both models (Table~\ref{S-tab:chi} of the supplementary material). This holds at equal pressure; at equal density the ordering is trivially reversed, because $\Delta=-0.1$ then sits at a much higher pressure. A plausible reading is that a more negative $\Delta$ lowers the cost of unlike contacts, so that the mixture accommodates more particles per unit volume at a given pressure and compresses more readily before the cages lock; our data are consistent with this picture.

\emph{Cage dynamics and arrest.} In the fluid states at $\bP\approx20$--$22$ the main-peak relaxation times of the PHS mixture are essentially the same, $\tau_\alpha=1.73$ and $1.58\,\tau_{\rm HS}$ for $\Delta=-0.1$ and $-0.2$, which is compatible with a slowing-down that is organized more by pressure than by density, although two state points cannot establish this. The onset of arrest points the same way: the Wendt--Abraham knee lies at $\bP\approx31.5$ ($\Delta=-0.1$) and $33.3$--$33.6$ ($\Delta=-0.2$), and the $\Delta=-0.2$ mixture still relaxes at $\bP\approx35$ whereas $\Delta=-0.1$ is arrested at $\bP\approx40$ on our time scale. The corresponding densities, about $1.28$ and $1.46$, differ by $14\%$, so that arrest is reached at a pressure window common to both but at a density that increases as $\Delta$ becomes more negative. In the glasses, which lie at similar pressures ($\bP=64.5$ and $60.3$), the $\Delta=-0.2$ cage is somewhat looser: the plateau MSD at $t=20\,\tau_{\rm HS}$ is $0.0124\sigma^2$ against $0.0098\sigma^2$ and the main-peak nonergodicity parameter is $0.878$ against $0.910$, although the density is higher. The $\Delta=-0.2$ glass is at a $7\%$ lower pressure, which may account for part of this difference.

\emph{Elastic response.} The elastic constants point the same way. At $\Delta=-0.2$ the shear plateau is lower ($G_{\rm plat}=196$ against $296\,\epsilon/\sigma^3$), the bulk modulus is lower ($1200$ against $1430$), and the sound speeds are lower, $c_T=11.3$ and $c_L=30.8$ against $14.7$ and $36.5\,\sigma/\tau_{\rm LJ}$, i.e.\ $2.3$ and $5.8$~km/s at $\Delta=-0.1$ and $1.8$ and $4.9$~km/s at $\Delta=-0.2$ for argon-like parameters. The bulk modulus decreases less than $G$, so $c_L/c_T$ and the Poisson ratio increase, from $2.48$ and $0.40$ to $2.73$ and $0.42$. The longitudinal speeds from the current spectra at the smallest wave vector are about $20\%$ above the isothermal estimates and the transverse ones agree with them, but both give the same ordering between the two glasses (Table~\ref{S-tab:currents} of the supplementary material). Given the uncertainties in Table~\ref{S-tab:sound} of the supplementary material, the differences in $c_L$ and $c_T$ exceed the statistical errors ($\approx4$ and $2.5$ standard deviations).

Taken together, these observations suggest that more negative non-additivity makes the mixture easier to pack and the resulting glass more compliant, with a larger compressibility and a lower shear stiffness at comparable pressure, while the pressure at which the dynamics arrests changes little.}

\begin{table}
\caption{\rnew{Comparison of the PHS mixtures at $\Delta=-0.1$ and $-0.2$ ($c=0.3$, $T^*=1.5$). Upper block: densities and dimensionless isothermal compressibility $\rho k_BT\kappa_T$ at equal pressure, from the PHS equations of state and, for NAHS, from the Monte Carlo tables of Ref.~\onlinecite{PPLS2026} (uncertainty $10$--$15\%$). Lower block: glasses (Sec.~\ref{sec:hsdyn} and the supplementary material); speeds in Lennard-Jones units, with argon-like values in km/s in parentheses.}}
\label{tab:nonadd}
\footnotesize
\begin{ruledtabular}
\setlength{\tabcolsep}{2pt}
\begin{tabular}{lcc}
 & $\Delta=-0.1$ & $\Delta=-0.2$\\
\hline
$\rho\sigma^3$ at $\bP=20,40,60$ & 1.188, 1.325, 1.367 & 1.319, 1.492, 1.545\\
$10^3\rho k_BT\kappa_T$, PHS & 10.4, 3.3, 1.6 & 15.3, 3.9, 1.9\\
$10^3\rho k_BT\kappa_T$, NAHS (MC) & 12.5, 3.0, 1.4 & 15.7, 4.1, 1.6\\
fluid $\tau_\alpha/\tau_{\rm HS}$ (main peak) & 1.73 ($\bP$=20) & 1.58 ($\bP$=22.15)\\
\hline
Glass $\rho\sigma^3$ ($\bP$) & 1.3707 (64.5) & 1.5442 (60.3)\\
MSD at $20\,\tau_{\rm HS}$ ($\sigma^2$) & 0.0098 & 0.0124\\
$f_Q$ (main peak) & 0.910 & 0.878\\
$G_{\rm plat}$, $K_T$ ($\epsilon/\sigma^3$) & 296, 1430 & 196, 1200\\
$c_T$ & 14.7 (2.3) & 11.3 (1.8)\\
$c_L$ & 36.5 (5.8) & 30.8 (4.9)\\
$c_L/c_T$, $\nu$ & 2.48, 0.40 & 2.73, 0.42\\
\end{tabular}
\end{ruledtabular}
\end{table}

\subsection{Softness dependence}\label{sec:soft}
\begin{figure}
\includegraphics[width=\columnwidth]{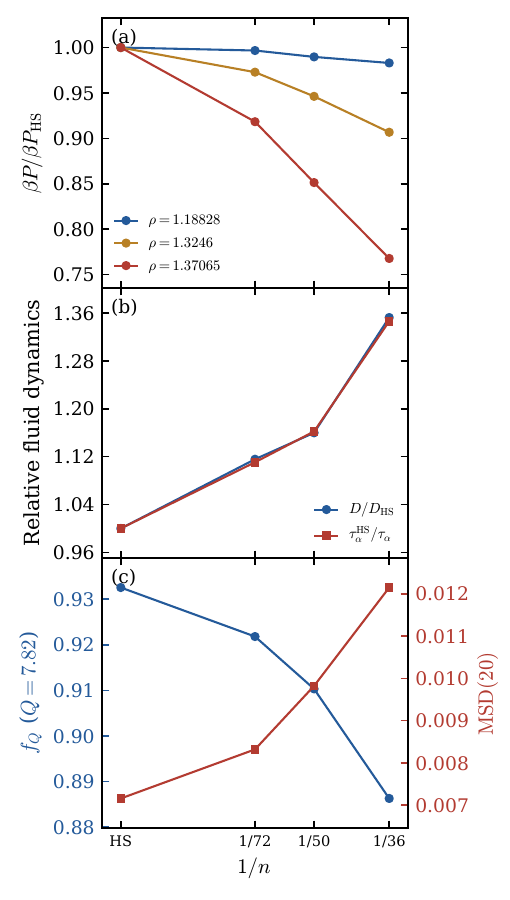}
\caption{Dependence on the repulsive exponent of Mie$(n,n-1)$ mixtures at $\Delta=-0.1$, $c=0.3$, plotted against $1/n$ with the NAHS (EDMD) limit at $1/n=0$. (a) Pressure relative to the NAHS value at three fixed densities. (b) Ratios of diffusion coefficients and of main-peak relaxation times to their NAHS values in the fluid at $\rho\sigma^3=1.18828$. (c) Main-peak nonergodicity parameter (left) and plateau MSD at $t=20\,\tau_{\rm HS}$ (right) in the glass at $\rho\sigma^3=1.3707$.}
\label{fig:soft}
\end{figure}

To separate softness effects from the specific choice $n=50$, we repeated the $\Delta=-0.1$ simulations with Mie$(n,n-1)$ repulsions for $n=36$ and $72$, cut and shifted at $r_{\rm min}=n\sigma_{ij}/(n-1)$. For each $n$ the temperature is fixed by matching the second virial coefficient to the hard-sphere value, calibrated so that $n=50$ gives $T^*=1.5$: $T^*(n)=1.5\,T^*_{B_2}(n)/T^*_{B_2}(50)$, which yields $T^*=1.508$ ($n=36$) and $1.494$ ($n=72$). Time is converted to $\tau_{\rm HS}$ with $\sqrt{T^*(n)}$.
The equation of state converges toward the NAHS reference with increasing $n$. At $\bP=20$ the $NPT$ densities are $1.1903$, $1.1883$, and $1.1868$ for $n=36$, 50, and 72, against $1.1876$ from the \giuseppe{data} of Ref.~\onlinecite{PPLS2026}; at $\bP=40$ they are $1.3302$, $1.3246$, and $1.3214$, against $1.3212$. Conversely, at fixed density $\rho\sigma^3=1.3246$ the pressure rises with $n$, $\bP=38.12$, $39.79$, and $40.90$, toward the EDMD value $42.05$.
The dynamics converge more slowly. At $\rho\sigma^3=1.1883$, where all systems relax, the main-peak relaxation time is $\tau_\alpha=1.49$, $1.73$, and $1.81\,\tau_{\rm HS}$ for $n=36$, 50, and 72, against $2.01\,\tau_{\rm HS}$ for the NAHS mixture, and $D/D_{\rm HS}=1.35$, $1.16$, and $1.11$. For $n\ge50$ the excess $D/D_{\rm HS}-1$ decreases approximately as $1/n$. At $\rho\sigma^3=1.3246$ none of the systems relaxes within $4800\,\tau_{\rm HS}$, but the plateau of $F_s(Q\sigma=7.82,t)$ at $t=10\,\tau_{\rm HS}$ rises ($0.805$, $0.823$, $0.832$) and the cage MSD falls ($0.0221$, $0.0197$, $0.0186\sigma^2$) monotonically with $n$ toward the NAHS values ($0.844$ and $0.0171\sigma^2$). The faster PHS dynamics is therefore a softness effect that vanishes in the hard-sphere limit, and at $n=50$ it amounts to about $15$--$20\%$ in the fluid. The glass shows the same convergence most clearly. Starting from the same configuration at $\rho\sigma^3=1.3707$, the pressure is $\bP=58.3$, $64.5$, and $69.7$ for $n=36$, 50, and 72, against $75.9$ for the NAHS mixture; the main-peak nonergodicity parameter is $f_Q=0.886$, $0.910$, and $0.922$, against $0.933$; and the plateau MSD at $t=20\,\tau_{\rm HS}$ is $0.0121$, $0.0098$, and $0.0083\sigma^2$, against $0.0072\sigma^2$. These trends are summarized in Fig.~\ref{fig:soft}.

\section{Discussion}\label{sec:disc}
The comparisons presented in this work separate quantities that the
PHS model transfers satisfactorily to
the NAHS mixture from those for which its performance
degrades. Thermodynamics and phase 
behavior transfer well when compared at equal pressure. The fluid
densities agree with the equations of state of
Ref.~\onlinecite{PPLS2026} within $0.8\%$, and direct coexistence
gives freezing pressures within $0.9\%$ for FCC and within $0.25\%$
for the B1 crystal of the strongly non-additive mixture. Because the
equation of state is steep, the same small density offsets appear as
larger pressure differences at fixed density: about $1\%$ at
$\bP\approx20$, $4$--$5\%$ at $\bP\approx40$, and \rev{$12$--$18\%$} in the
glasses. The comparison variable must therefore be chosen consciously;
at equal density the PHS mixture is effectively at a slightly lower
pressure than the NAHS mixture. 

The structure beyond contact is also transferable. In the ergodic
fluids the differences are small but significant, with the main peak
of $S_{NN}(Q)$ $0.5$--$1.3\%$ lower in PHS, while $S_{cc}(Q)$ agrees
within the statistical noise; in the glasses, $S_{NN}(Q)$,
$S_{cc}(Q)$, and the partial $g_{ij}(r)$ beyond the first peak agree
within the scatter between independently prepared samples. The
\santi{compositional order} that controls glass formation in these mixtures is
therefore captured. As could be expected, what differs is the contact
region: the soft 
repulsion lowers the contact values of $g_{ij}(r)$ by about a quarter
in the glasses, and through the virial route this accounts for the
lower PHS pressure at equal density. The independently prepared
glasses also show that sample-to-sample variation is substantial at
these densities.  

Slow dynamics transfers approximately. In the fluid, PHS relaxation
is faster than NAHS relaxation by a single factor, $1.16$ at
$\Delta=-0.1$ and $1.22$ at $\Delta=-0.2$, which is the same for
$\tau_\alpha$ at all wave numbers and for $D$. A constant rescaling of
time therefore maps PHS fluid dynamics onto NAHS dynamics. This
factor grows on approaching arrest, to about $2$ at $\Delta=-0.2$,
$\bP\approx35$, so no single time unit maps the whole
slowing-down. The growth parallels the softer PHS cage: the plateau
MSD is $10$--$36\%$ larger and the nonergodicity parameters are lower
than for hard cores at the same density, so that particles explore
more of the cage and escape sooner. The softness series supports this
interpretation. With increasing exponent, the pressure, $D$,
$\tau_\alpha$, $f_Q$, and the cage size all approach the NAHS values
monotonically, and for $n\ge50$ the excess diffusion decreases roughly
as $1/n$. The faster dynamics is therefore a controlled softness
effect rather than a failure of the mapping, and it can be reduced by
increasing $n$ at the cost of a smaller time step. Comparing at equal
pressure rather than equal density removes about a quarter to a third
of the acceleration in the fluid (Sec.~\ref{sec:hsdyn}), so most of it
is intrinsic to the softer potential. Near arrest the picture is less
clear-cut. At $\bP\approx40$--$42$, where the equal-density pressure
difference is largest, the equal-pressure comparison removes most of
the acceleration, whereas at $\Delta=-0.2$, $\bP\approx35$ a factor of
about $2$ or more survives. The near-arrest factor is therefore partly
a state-point effect and is uncertain by the aging and trajectory
scatter of the hard-core runs. 

For the glass-forming problem of Ref.~\onlinecite{PPLS2026}, these
results mean that the PHS model can be used to locate phase
boundaries, to characterize local and \santi{compositional order}, and to map the
glass-forming window, but that relaxation times near arrest are
underestimated, here by about a factor of two or more at
$\bP\approx35$, at equal density and at equal pressure alike. At
$\Delta=-0.1$ both models are already arrested on our time scale at
$\bP\approx40$, whereas at $\Delta=-0.2$ the mixture still relaxes at
$\bP\approx35$; since the PHS mixture relaxes faster at equal
density, arrest densities obtained from PHS dynamics are upper bounds
for the hard-core mixture. The Wendt--Abraham ratio, $R_{\rm
  WA}=g_{NN}(r_{\min})/g_{NN}(r_{\max})$, between the first minimum
and the first maximum of the number--number distribution function
$g_{NN}=x_1^2g_{11}+2x_1x_2g_{12}+x_2^2g_{22}$, gives a structural
view of the same trend (Fig.~\ref{S-fig:wa} of the supplementary
material).\cite{Wendt1978} Ref.~\onlinecite{PPLS2026} found a knee in
$R_{\rm WA}(\bP)$ near $\bP=35$ for $\Delta=-0.2$. Fitting two
straight lines to $R_{\rm WA}$ over $20\le\bP\le75$, we locate the
knee  at $\bP=33.3$--$33.6$ for PHS, a
difference comparable to the spacing of the PHS states. At
$\Delta=-0.1$ the PHS knee lies at $\bP\approx 31.5$, close to the
onset of arrest seen in the dynamics. The PHS values of $R_{\rm WA}$
are slightly higher than the hard-core ones (for example $0.128$
against $\approx 0.12$ at $\bP\approx 35$ for $\Delta=-0.2$), because the softer repulsion
lowers the contact peak while the first minimum is the same in both
models. The PHS 
mixture thus reproduces the structural signature of arrest at nearly
the same pressure, although, like its dynamics, its absolute $R_{\rm
  WA}$ is slightly shifted by the softness of the repulsion. 

\section{Conclusions}\label{sec:concl}
We have tested a Mie(50,49) pseudo-hard-sphere model at $T^*=1.5$ as a
representation of symmetric, negatively non-additive hard-sphere
mixtures. The model reproduces the NAHS equations of state within
$0.8\%$ in density and the freezing pressures within $0.9\%$ (FCC) and
$0.25\%$ (B1), and its fluid and glass structures agree closely with
those of the hard-core mixture beyond contact. Its dynamics is
somewhat faster
than that of the hard-core mixture at equal density, by a
wave-number-independent factor of $1.16$--$1.22$ in the fluid that
grows to about $2$ near dynamical arrest, together with a softer cage
and lower nonergodicity parameters. At equal pressure the fluid factor
drops to $1.12$--$1.18$; near arrest the equal-pressure difference
falls to the level of the trajectory scatter at the highest pressures
but persists at $\Delta=-0.2$, $\bP\approx35$. Varying the repulsive
exponent shows that all these differences decrease monotonically
toward the hard-sphere limit. Pseudo-hard-sphere models of this kind
are thus quantitative for the thermodynamics, phase behavior, and
structure of non-additive hard-sphere mixtures, and semi-quantitative
for their slow dynamics, for which the near-arrest acceleration should
be taken into account or removed with steeper repulsions. 

\rnew{The elastic response of these glasses also invites a comparison with real amorphous solids. With argon-like parameters, the PHS glasses have longitudinal and transverse sound speeds $c_L\approx5.8$ and $c_T\approx2.3$~km/s at $\Delta=-0.1$ ($4.9$ and $1.8$~km/s at $\Delta=-0.2$), a shear modulus $G\approx12$ ($8$)~GPa, and an isothermal bulk modulus $K_T\approx60$ ($50$)~GPa (Table~\ref{tab:nonadd} and Supplementary Material). These speeds fall in the range measured for metallic glasses, $c_L\approx4$--$6$ and $c_T\approx2$--$2.7$~km/s.\cite{Wang2012} More significantly, the unit-independent ratios $c_L/c_T\approx2.5$--$2.7$ correspond to Poisson ratios $\nu\approx0.40$--$0.42$ and to $K/G\approx5$--$6$, values also characteristic of metallic glasses ($\nu\approx0.3$--$0.42$) and well above those of oxide glasses ($\nu\approx0.2$), and associated in metallic glasses with intrinsic plasticity.\cite{Lewandowski2005,Wang2012} Dense non-additive hard-sphere glasses, dominated by the packing of nearly hard cores of two sizes with a preference for unlike contacts, may thus provide a minimal model for the high Poisson ratio of metallic glasses, although these estimates use isothermal compressibilities and a finite-window shear plateau and are only indicative.}

\begin{acknowledgments}
E.L. acknowledges the financial support from Grant No.\ PID2023-151751NB-I00 funded by MICIU/AEI/10.13039/501100011033 and by ``ERDF A way of making Europe.''  We also acknowledge the Galicia Supercomputing Center
(CESGA) for the generous access to their computer facilities. \giuseppe{G.P.\ would like to acknowledge the South African Centre for High Performance Computing (CHPC) for granting access to computational resources under the allocation MATS0887.}
\end{acknowledgments}
\section*{Supplementary material}
The supplementary material contains the structure and dynamics of the $\Delta=-0.1$ glass, including the partial pair distributions near contact for both glasses; the stress relaxation and its Zwanzig--Mountain check; the elastic constants and sound speeds of the PHS glasses from the equation of state, the shear modulus, and the current spectra; the vibrational spectra; the bond-orientational order of the B1 solid; and the Wendt--Abraham ratio.
\section*{Data availability}
The data that support the findings of this study are available from the authors upon request.
\bibliography{refs}

\begin{thebibliography}{36}%
\makeatletter
\providecommand \@ifxundefined [1]{%
 \@ifx{#1\undefined}
}%
\providecommand \@ifnum [1]{%
 \ifnum #1\expandafter \@firstoftwo
 \else \expandafter \@secondoftwo
 \fi
}%
\providecommand \@ifx [1]{%
 \ifx #1\expandafter \@firstoftwo
 \else \expandafter \@secondoftwo
 \fi
}%
\providecommand \natexlab [1]{#1}%
\providecommand \enquote  [1]{``#1''}%
\providecommand \bibnamefont  [1]{#1}%
\providecommand \bibfnamefont [1]{#1}%
\providecommand \citenamefont [1]{#1}%
\providecommand \href@noop [0]{\@secondoftwo}%
\providecommand \href [0]{\begingroup \@sanitize@url \@href}%
\providecommand \@href[1]{\@@startlink{#1}\@@href}%
\providecommand \@@href[1]{\endgroup#1\@@endlink}%
\providecommand \@sanitize@url [0]{\catcode `\\12\catcode `\$12\catcode
  `\&12\catcode `\#12\catcode `\^12\catcode `\_12\catcode `\%12\relax}%
\providecommand \@@startlink[1]{}%
\providecommand \@@endlink[0]{}%
\providecommand \url  [0]{\begingroup\@sanitize@url \@url }%
\providecommand \@url [1]{\endgroup\@href {#1}{\urlprefix }}%
\providecommand \urlprefix  [0]{URL }%
\providecommand \Eprint [0]{\href }%
\providecommand \doibase [0]{https://doi.org/}%
\providecommand \selectlanguage [0]{\@gobble}%
\providecommand \bibinfo  [0]{\@secondoftwo}%
\providecommand \bibfield  [0]{\@secondoftwo}%
\providecommand \translation [1]{[#1]}%
\providecommand \BibitemOpen [0]{}%
\providecommand \bibitemStop [0]{}%
\providecommand \bibitemNoStop [0]{.\EOS\space}%
\providecommand \EOS [0]{\spacefactor3000\relax}%
\providecommand \BibitemShut  [1]{\csname bibitem#1\endcsname}%
\let\auto@bib@innerbib\@empty
\bibitem [{\citenamefont {Schroers}(2013)}]{Schroers2013}%
  \BibitemOpen
  \bibfield  {author} {\bibinfo {author} {\bibfnamefont {J.}~\bibnamefont
  {Schroers}},\ }\bibfield  {title} {\enquote {\bibinfo {title} {Bulk metallic
  glasses},}\ }\href {https://doi.org/10.1063/PT.3.1885} {\bibfield  {journal}
  {\bibinfo  {journal} {Phys. Today}\ }\textbf {\bibinfo {volume} {66}},\
  \bibinfo {pages} {32--37} (\bibinfo {year} {2013})}\BibitemShut {NoStop}%
\bibitem [{\citenamefont {Turnbull}(1969)}]{Turnbull1969}%
  \BibitemOpen
  \bibfield  {author} {\bibinfo {author} {\bibfnamefont {D.}~\bibnamefont
  {Turnbull}},\ }\bibfield  {title} {\enquote {\bibinfo {title} {Under what
  conditions can a glass be formed?}}\ }\href
  {https://doi.org/10.1080/00107516908204405} {\bibfield  {journal} {\bibinfo
  {journal} {Contemp. Phys.}\ }\textbf {\bibinfo {volume} {10}},\ \bibinfo
  {pages} {473--488} (\bibinfo {year} {1969})}\BibitemShut {NoStop}%
\bibitem [{\citenamefont {Yadav}\ \emph {et~al.}(2025)\citenamefont {Yadav},
  \citenamefont {Liu}, \citenamefont {Wu}, \citenamefont {Yadav},\ and\
  \citenamefont {Zheng}}]{Yadav2025}%
  \BibitemOpen
  \bibfield  {author} {\bibinfo {author} {\bibfnamefont {N.}~\bibnamefont
  {Yadav}}, \bibinfo {author} {\bibfnamefont {M.}~\bibnamefont {Liu}}, \bibinfo
  {author} {\bibfnamefont {Y.}~\bibnamefont {Wu}}, \bibinfo {author}
  {\bibfnamefont {A.}~\bibnamefont {Yadav}},\ and\ \bibinfo {author}
  {\bibfnamefont {H.}~\bibnamefont {Zheng}},\ }\bibfield  {title} {\enquote
  {\bibinfo {title} {Self-organization of photonic structures in colloidal
  crystals in the {AI} era},}\ }\href {https://doi.org/10.1039/D5NA00449G}
  {\bibfield  {journal} {\bibinfo  {journal} {Nanoscale Adv.}\ }\textbf
  {\bibinfo {volume} {7}},\ \bibinfo {pages} {6373--6393} (\bibinfo {year}
  {2025})}\BibitemShut {NoStop}%
\bibitem [{\citenamefont {Hu}\ \emph {et~al.}(2021)\citenamefont {Hu},
  \citenamefont {Zhang}, \citenamefont {Yang}, \citenamefont {Ma},\ and\
  \citenamefont {Huang}}]{Hu2021}%
  \BibitemOpen
  \bibfield  {author} {\bibinfo {author} {\bibfnamefont {Y.}~\bibnamefont
  {Hu}}, \bibinfo {author} {\bibfnamefont {Y.}~\bibnamefont {Zhang}}, \bibinfo
  {author} {\bibfnamefont {D.}~\bibnamefont {Yang}}, \bibinfo {author}
  {\bibfnamefont {D.}~\bibnamefont {Ma}},\ and\ \bibinfo {author}
  {\bibfnamefont {S.}~\bibnamefont {Huang}},\ }\bibfield  {title} {\enquote
  {\bibinfo {title} {Self-assembly of colloidal particles into amorphous
  photonic crystals},}\ }\href {https://doi.org/10.1039/D1MA00477H} {\bibfield
  {journal} {\bibinfo  {journal} {Mater. Adv.}\ }\textbf {\bibinfo {volume}
  {2}},\ \bibinfo {pages} {6499--6518} (\bibinfo {year} {2021})}\BibitemShut
  {NoStop}%
\bibitem [{\citenamefont {Kob}\ and\ \citenamefont {Andersen}(1994)}]{Kob1994}%
  \BibitemOpen
  \bibfield  {author} {\bibinfo {author} {\bibfnamefont {W.}~\bibnamefont
  {Kob}}\ and\ \bibinfo {author} {\bibfnamefont {H.~C.}\ \bibnamefont
  {Andersen}},\ }\bibfield  {title} {\enquote {\bibinfo {title} {Scaling
  behavior in the {$\beta$}-relaxation regime of a supercooled
  {L}ennard-{J}ones mixture},}\ }\href
  {https://doi.org/10.1103/PhysRevLett.73.1376} {\bibfield  {journal} {\bibinfo
   {journal} {Phys. Rev. Lett.}\ }\textbf {\bibinfo {volume} {73}},\ \bibinfo
  {pages} {1376--1379} (\bibinfo {year} {1994})}\BibitemShut {NoStop}%
\bibitem [{\citenamefont {Pedersen}, \citenamefont {Schr{\o}der},\ and\
  \citenamefont {Dyre}(2018)}]{Pedersen2018}%
  \BibitemOpen
  \bibfield  {author} {\bibinfo {author} {\bibfnamefont {U.~R.}\ \bibnamefont
  {Pedersen}}, \bibinfo {author} {\bibfnamefont {T.~B.}\ \bibnamefont
  {Schr{\o}der}},\ and\ \bibinfo {author} {\bibfnamefont {J.~C.}\ \bibnamefont
  {Dyre}},\ }\bibfield  {title} {\enquote {\bibinfo {title} {Phase diagram of
  {Kob-Andersen}-type binary {Lennard-Jones} mixtures},}\ }\href
  {https://doi.org/10.1103/PhysRevLett.120.165501} {\bibfield  {journal}
  {\bibinfo  {journal} {Phys. Rev. Lett.}\ }\textbf {\bibinfo {volume} {120}},\
  \bibinfo {pages} {165501} (\bibinfo {year} {2018})}\BibitemShut {NoStop}%
\bibitem [{\citenamefont {Zhang}\ \emph {et~al.}(2015)\citenamefont {Zhang},
  \citenamefont {Fan}, \citenamefont {Liu}, \citenamefont {Schroers},
  \citenamefont {Shattuck},\ and\ \citenamefont {O'Hern}}]{Zhang2015}%
  \BibitemOpen
  \bibfield  {author} {\bibinfo {author} {\bibfnamefont {J.}~\bibnamefont
  {Zhang}}, \bibinfo {author} {\bibfnamefont {Y.}~\bibnamefont {Fan}}, \bibinfo
  {author} {\bibfnamefont {J.}~\bibnamefont {Liu}}, \bibinfo {author}
  {\bibfnamefont {J.}~\bibnamefont {Schroers}}, \bibinfo {author}
  {\bibfnamefont {M.~D.}\ \bibnamefont {Shattuck}},\ and\ \bibinfo {author}
  {\bibfnamefont {C.~S.}\ \bibnamefont {O'Hern}},\ }\bibfield  {title}
  {\enquote {\bibinfo {title} {Beyond packing of hard spheres: The effects of
  core softness, non-additivity, intermediate-range repulsion, and many-body
  interactions on the glass-forming ability of bulk metallic glasses},}\ }\href
  {https://doi.org/10.1063/1.4935002} {\bibfield  {journal} {\bibinfo
  {journal} {J. Chem. Phys.}\ }\textbf {\bibinfo {volume} {143}},\ \bibinfo
  {pages} {184502} (\bibinfo {year} {2015})}\BibitemShut {NoStop}%
\bibitem [{\citenamefont {Pusey}\ and\ \citenamefont {van
  Megen}(1986)}]{Pusey1986}%
  \BibitemOpen
  \bibfield  {author} {\bibinfo {author} {\bibfnamefont {P.~N.}\ \bibnamefont
  {Pusey}}\ and\ \bibinfo {author} {\bibfnamefont {W.}~\bibnamefont {van
  Megen}},\ }\bibfield  {title} {\enquote {\bibinfo {title} {Phase behaviour of
  concentrated suspensions of nearly hard colloidal spheres},}\ }\href
  {https://doi.org/10.1038/320340a0} {\bibfield  {journal} {\bibinfo  {journal}
  {Nature}\ }\textbf {\bibinfo {volume} {320}},\ \bibinfo {pages} {340--342}
  (\bibinfo {year} {1986})}\BibitemShut {NoStop}%
\bibitem [{\citenamefont {Zaccarelli}\ \emph {et~al.}(2009)\citenamefont
  {Zaccarelli}, \citenamefont {Valeriani}, \citenamefont {Sanz}, \citenamefont
  {Poon}, \citenamefont {Cates},\ and\ \citenamefont {Pusey}}]{Zaccarelli2009}%
  \BibitemOpen
  \bibfield  {author} {\bibinfo {author} {\bibfnamefont {E.}~\bibnamefont
  {Zaccarelli}}, \bibinfo {author} {\bibfnamefont {C.}~\bibnamefont
  {Valeriani}}, \bibinfo {author} {\bibfnamefont {E.}~\bibnamefont {Sanz}},
  \bibinfo {author} {\bibfnamefont {W.~C.~K.}\ \bibnamefont {Poon}}, \bibinfo
  {author} {\bibfnamefont {M.~E.}\ \bibnamefont {Cates}},\ and\ \bibinfo
  {author} {\bibfnamefont {P.~N.}\ \bibnamefont {Pusey}},\ }\bibfield  {title}
  {\enquote {\bibinfo {title} {Crystallization of hard-sphere glasses},}\
  }\href {https://doi.org/10.1103/PhysRevLett.103.135704} {\bibfield  {journal}
  {\bibinfo  {journal} {Phys. Rev. Lett.}\ }\textbf {\bibinfo {volume} {103}},\
  \bibinfo {pages} {135704} (\bibinfo {year} {2009})}\BibitemShut {NoStop}%
\bibitem [{\citenamefont {Berthier}\ and\ \citenamefont
  {Biroli}(2011)}]{BerthierBiroli2011}%
  \BibitemOpen
  \bibfield  {author} {\bibinfo {author} {\bibfnamefont {L.}~\bibnamefont
  {Berthier}}\ and\ \bibinfo {author} {\bibfnamefont {G.}~\bibnamefont
  {Biroli}},\ }\bibfield  {title} {\enquote {\bibinfo {title} {Theoretical
  perspective on the glass transition and amorphous materials},}\ }\href
  {https://doi.org/10.1103/RevModPhys.83.587} {\bibfield  {journal} {\bibinfo
  {journal} {Rev. Mod. Phys.}\ }\textbf {\bibinfo {volume} {83}},\ \bibinfo
  {pages} {587--645} (\bibinfo {year} {2011})}\BibitemShut {NoStop}%
\bibitem [{\citenamefont {Royall}\ and\ \citenamefont
  {Williams}(2015)}]{Royall2015}%
  \BibitemOpen
  \bibfield  {author} {\bibinfo {author} {\bibfnamefont {C.~P.}\ \bibnamefont
  {Royall}}\ and\ \bibinfo {author} {\bibfnamefont {S.~R.}\ \bibnamefont
  {Williams}},\ }\bibfield  {title} {\enquote {\bibinfo {title} {The role of
  local structure in dynamical arrest},}\ }\href
  {https://doi.org/10.1016/j.physrep.2014.11.004} {\bibfield  {journal}
  {\bibinfo  {journal} {Phys. Rep.}\ }\textbf {\bibinfo {volume} {560}},\
  \bibinfo {pages} {1--75} (\bibinfo {year} {2015})}\BibitemShut {NoStop}%
\bibitem [{\citenamefont {L{\'a}zaro-L{\'a}zaro}\ \emph
  {et~al.}(2019)\citenamefont {L{\'a}zaro-L{\'a}zaro}, \citenamefont
  {Perera-Burgos}, \citenamefont {Laermann}, \citenamefont {Sentjabrskaja},
  \citenamefont {P{\'e}rez-{\'A}ngel}, \citenamefont {Laurati}, \citenamefont
  {Egelhaaf}, \citenamefont {Medina-Noyola}, \citenamefont {Voigtmann},
  \citenamefont {Casta{\~n}eda-Priego},\ and\ \citenamefont
  {Elizondo-Aguilera}}]{Lazaro2019}%
  \BibitemOpen
  \bibfield  {author} {\bibinfo {author} {\bibfnamefont {E.}~\bibnamefont
  {L{\'a}zaro-L{\'a}zaro}}, \bibinfo {author} {\bibfnamefont {J.~A.}\
  \bibnamefont {Perera-Burgos}}, \bibinfo {author} {\bibfnamefont
  {P.}~\bibnamefont {Laermann}}, \bibinfo {author} {\bibfnamefont
  {T.}~\bibnamefont {Sentjabrskaja}}, \bibinfo {author} {\bibfnamefont
  {G.}~\bibnamefont {P{\'e}rez-{\'A}ngel}}, \bibinfo {author} {\bibfnamefont
  {M.}~\bibnamefont {Laurati}}, \bibinfo {author} {\bibfnamefont {S.~U.}\
  \bibnamefont {Egelhaaf}}, \bibinfo {author} {\bibfnamefont {M.}~\bibnamefont
  {Medina-Noyola}}, \bibinfo {author} {\bibfnamefont {T.}~\bibnamefont
  {Voigtmann}}, \bibinfo {author} {\bibfnamefont {R.}~\bibnamefont
  {Casta{\~n}eda-Priego}},\ and\ \bibinfo {author} {\bibfnamefont {L.~F.}\
  \bibnamefont {Elizondo-Aguilera}},\ }\bibfield  {title} {\enquote {\bibinfo
  {title} {Glassy dynamics in asymmetric binary mixtures of hard spheres},}\
  }\href {https://doi.org/10.1103/PhysRevE.99.042603} {\bibfield  {journal}
  {\bibinfo  {journal} {Phys. Rev. E}\ }\textbf {\bibinfo {volume} {99}},\
  \bibinfo {pages} {042603} (\bibinfo {year} {2019})}\BibitemShut {NoStop}%
\bibitem [{\citenamefont {Punnathanam}\ and\ \citenamefont
  {Monson}(2006)}]{Punnathanam2006}%
  \BibitemOpen
  \bibfield  {author} {\bibinfo {author} {\bibfnamefont {S.}~\bibnamefont
  {Punnathanam}}\ and\ \bibinfo {author} {\bibfnamefont {P.~A.}\ \bibnamefont
  {Monson}},\ }\bibfield  {title} {\enquote {\bibinfo {title} {Crystal
  nucleation in binary hard sphere mixtures: A {M}onte {C}arlo simulation
  study},}\ }\href {https://doi.org/10.1063/1.2208998} {\bibfield  {journal}
  {\bibinfo  {journal} {J. Chem. Phys.}\ }\textbf {\bibinfo {volume} {125}},\
  \bibinfo {pages} {024508} (\bibinfo {year} {2006})}\BibitemShut {NoStop}%
\bibitem [{\citenamefont {Prestipino}\ \emph {et~al.}(2026)\citenamefont
  {Prestipino}, \citenamefont {Pellicane}, \citenamefont {Lomba},\ and\
  \citenamefont {Saija}}]{PPLS2026}%
  \BibitemOpen
  \bibfield  {author} {\bibinfo {author} {\bibfnamefont {S.}~\bibnamefont
  {Prestipino}}, \bibinfo {author} {\bibfnamefont {G.}~\bibnamefont
  {Pellicane}}, \bibinfo {author} {\bibfnamefont {E.}~\bibnamefont {Lomba}},\
  and\ \bibinfo {author} {\bibfnamefont {F.}~\bibnamefont {Saija}},\ }\href
  {https://doi.org/10.48550/arXiv.2611.20944} {\enquote {\bibinfo {title}
  {Polymorphism and crystal suppression in symmetric mixtures with entropic
  interactions},}\ } (\bibinfo {year} {2026}),\ \bibinfo {note}
  {arXiv:2611.20944}\BibitemShut {NoStop}%
\bibitem [{\citenamefont {Bannerman}, \citenamefont {Sargant},\ and\
  \citenamefont {Lue}(2011)}]{Bannerman2011}%
  \BibitemOpen
  \bibfield  {author} {\bibinfo {author} {\bibfnamefont {M.~N.}\ \bibnamefont
  {Bannerman}}, \bibinfo {author} {\bibfnamefont {R.}~\bibnamefont {Sargant}},\
  and\ \bibinfo {author} {\bibfnamefont {L.}~\bibnamefont {Lue}},\ }\bibfield
  {title} {\enquote {\bibinfo {title} {{DynamO}: a free {$\mathcal{O}(N)$}
  general event-driven molecular dynamics simulator},}\ }\href
  {https://doi.org/10.1002/jcc.21915} {\bibfield  {journal} {\bibinfo
  {journal} {J. Comput. Chem.}\ }\textbf {\bibinfo {volume} {32}},\ \bibinfo
  {pages} {3329--3338} (\bibinfo {year} {2011})}\BibitemShut {NoStop}%
\bibitem [{\citenamefont {Jover}\ \emph {et~al.}(2012)\citenamefont {Jover},
  \citenamefont {Haslam}, \citenamefont {Galindo}, \citenamefont {Jackson},\
  and\ \citenamefont {M\"uller}}]{Jover2012}%
  \BibitemOpen
  \bibfield  {author} {\bibinfo {author} {\bibfnamefont {J.}~\bibnamefont
  {Jover}}, \bibinfo {author} {\bibfnamefont {A.~J.}\ \bibnamefont {Haslam}},
  \bibinfo {author} {\bibfnamefont {A.}~\bibnamefont {Galindo}}, \bibinfo
  {author} {\bibfnamefont {G.}~\bibnamefont {Jackson}},\ and\ \bibinfo {author}
  {\bibfnamefont {E.~A.}\ \bibnamefont {M\"uller}},\ }\bibfield  {title}
  {\enquote {\bibinfo {title} {Pseudo hard-sphere potential for use in
  continuous molecular-dynamics simulation of spherical and chain molecules},}\
  }\href {https://doi.org/10.1063/1.4754275} {\bibfield  {journal} {\bibinfo
  {journal} {J. Chem. Phys.}\ }\textbf {\bibinfo {volume} {137}},\ \bibinfo
  {pages} {144505} (\bibinfo {year} {2012})}\BibitemShut {NoStop}%
\bibitem [{\citenamefont {Espinosa}\ \emph {et~al.}(2013)\citenamefont
  {Espinosa}, \citenamefont {Sanz}, \citenamefont {Valeriani},\ and\
  \citenamefont {Vega}}]{Espinosa2013}%
  \BibitemOpen
  \bibfield  {author} {\bibinfo {author} {\bibfnamefont {J.~R.}\ \bibnamefont
  {Espinosa}}, \bibinfo {author} {\bibfnamefont {E.}~\bibnamefont {Sanz}},
  \bibinfo {author} {\bibfnamefont {C.}~\bibnamefont {Valeriani}},\ and\
  \bibinfo {author} {\bibfnamefont {C.}~\bibnamefont {Vega}},\ }\bibfield
  {title} {\enquote {\bibinfo {title} {On fluid-solid direct coexistence
  simulations: The pseudo-hard sphere model},}\ }\href
  {https://doi.org/10.1063/1.4823499} {\bibfield  {journal} {\bibinfo
  {journal} {J. Chem. Phys.}\ }\textbf {\bibinfo {volume} {139}},\ \bibinfo
  {pages} {144502} (\bibinfo {year} {2013})}\BibitemShut {NoStop}%
\bibitem [{\citenamefont {B{\'a}ez}\ \emph {et~al.}(2018)\citenamefont
  {B{\'a}ez}, \citenamefont {Torres-Carbajal}, \citenamefont
  {Casta{\~n}eda-Priego}, \citenamefont {Villada-Balbuena}, \citenamefont
  {M{\'e}ndez-Alcaraz},\ and\ \citenamefont {Herrera-Velarde}}]{Baez2018}%
  \BibitemOpen
  \bibfield  {author} {\bibinfo {author} {\bibfnamefont {C.~A.}\ \bibnamefont
  {B{\'a}ez}}, \bibinfo {author} {\bibfnamefont {A.}~\bibnamefont
  {Torres-Carbajal}}, \bibinfo {author} {\bibfnamefont {R.}~\bibnamefont
  {Casta{\~n}eda-Priego}}, \bibinfo {author} {\bibfnamefont {A.}~\bibnamefont
  {Villada-Balbuena}}, \bibinfo {author} {\bibfnamefont {J.~M.}\ \bibnamefont
  {M{\'e}ndez-Alcaraz}},\ and\ \bibinfo {author} {\bibfnamefont
  {S.}~\bibnamefont {Herrera-Velarde}},\ }\bibfield  {title} {\enquote
  {\bibinfo {title} {Using the second virial coefficient as physical criterion
  to map the hard-sphere potential onto a continuous potential},}\ }\href
  {https://doi.org/10.1063/1.5049568} {\bibfield  {journal} {\bibinfo
  {journal} {J. Chem. Phys.}\ }\textbf {\bibinfo {volume} {149}},\ \bibinfo
  {pages} {164907} (\bibinfo {year} {2018})}\BibitemShut {NoStop}%
\bibitem [{\citenamefont {Mart{\'i}nez-Rivera}\ \emph
  {et~al.}(2023)\citenamefont {Mart{\'i}nez-Rivera}, \citenamefont
  {Villada-Balbuena}, \citenamefont {Sandoval-Puentes}, \citenamefont
  {Egelhaaf}, \citenamefont {M{\'e}ndez-Alcaraz}, \citenamefont
  {Casta{\~n}eda-Priego},\ and\ \citenamefont
  {Escobedo-S{\'a}nchez}}]{MartinezRivera2023}%
  \BibitemOpen
  \bibfield  {author} {\bibinfo {author} {\bibfnamefont {J.}~\bibnamefont
  {Mart{\'i}nez-Rivera}}, \bibinfo {author} {\bibfnamefont {A.}~\bibnamefont
  {Villada-Balbuena}}, \bibinfo {author} {\bibfnamefont {M.~A.}\ \bibnamefont
  {Sandoval-Puentes}}, \bibinfo {author} {\bibfnamefont {S.~U.}\ \bibnamefont
  {Egelhaaf}}, \bibinfo {author} {\bibfnamefont {J.~M.}\ \bibnamefont
  {M{\'e}ndez-Alcaraz}}, \bibinfo {author} {\bibfnamefont {R.}~\bibnamefont
  {Casta{\~n}eda-Priego}},\ and\ \bibinfo {author} {\bibfnamefont {M.~A.}\
  \bibnamefont {Escobedo-S{\'a}nchez}},\ }\bibfield  {title} {\enquote
  {\bibinfo {title} {Modeling the structure and thermodynamics of
  multicomponent and polydisperse hard-sphere dispersions with continuous
  potentials},}\ }\href {https://doi.org/10.1063/5.0168098} {\bibfield
  {journal} {\bibinfo  {journal} {J. Chem. Phys.}\ }\textbf {\bibinfo {volume}
  {159}},\ \bibinfo {pages} {194110} (\bibinfo {year} {2023})}\BibitemShut
  {NoStop}%
\bibitem [{\citenamefont {de~los Santos-L{\'o}pez}\ \emph
  {et~al.}(2022)\citenamefont {de~los Santos-L{\'o}pez}, \citenamefont
  {P{\'e}rez-{\'A}ngel}, \citenamefont {Casta{\~n}eda-Priego},\ and\
  \citenamefont {M{\'e}ndez-Alcaraz}}]{SantosLopez2022}%
  \BibitemOpen
  \bibfield  {author} {\bibinfo {author} {\bibfnamefont {N.~M.}\ \bibnamefont
  {de~los Santos-L{\'o}pez}}, \bibinfo {author} {\bibfnamefont
  {G.}~\bibnamefont {P{\'e}rez-{\'A}ngel}}, \bibinfo {author} {\bibfnamefont
  {R.}~\bibnamefont {Casta{\~n}eda-Priego}},\ and\ \bibinfo {author}
  {\bibfnamefont {J.~M.}\ \bibnamefont {M{\'e}ndez-Alcaraz}},\ }\bibfield
  {title} {\enquote {\bibinfo {title} {Determining depletion interactions by
  contracting forces},}\ }\href {https://doi.org/10.1063/5.0099919} {\bibfield
  {journal} {\bibinfo  {journal} {J. Chem. Phys.}\ }\textbf {\bibinfo {volume}
  {157}},\ \bibinfo {pages} {074903} (\bibinfo {year} {2022})}\BibitemShut
  {NoStop}%
\bibitem [{\citenamefont {Bedolla-Montiel}, \citenamefont
  {Casta{\~n}eda-Cerd{\'a}n},\ and\ \citenamefont
  {Casta{\~n}eda-Priego}(2025)}]{BedollaMontiel2025}%
  \BibitemOpen
  \bibfield  {author} {\bibinfo {author} {\bibfnamefont {E.~A.}\ \bibnamefont
  {Bedolla-Montiel}}, \bibinfo {author} {\bibfnamefont {R.~A.}\ \bibnamefont
  {Casta{\~n}eda-Cerd{\'a}n}},\ and\ \bibinfo {author} {\bibfnamefont
  {R.}~\bibnamefont {Casta{\~n}eda-Priego}},\ }\bibfield  {title} {\enquote
  {\bibinfo {title} {Phase diagram of the hard-sphere potential model in three
  and four dimensions using a pseudo-hard-sphere potential},}\ }\href
  {https://doi.org/10.1063/5.0264356} {\bibfield  {journal} {\bibinfo
  {journal} {J. Chem. Phys.}\ }\textbf {\bibinfo {volume} {162}},\ \bibinfo
  {pages} {164504} (\bibinfo {year} {2025})}\BibitemShut {NoStop}%
\bibitem [{\citenamefont {Bedolla-Montiel}\ \emph {et~al.}(2024)\citenamefont
  {Bedolla-Montiel}, \citenamefont {Lange}, \citenamefont {P{\'e}rez~de
  Alba~Ort{\'i}z},\ and\ \citenamefont {Dijkstra}}]{BedollaMontiel2024}%
  \BibitemOpen
  \bibfield  {author} {\bibinfo {author} {\bibfnamefont {E.~A.}\ \bibnamefont
  {Bedolla-Montiel}}, \bibinfo {author} {\bibfnamefont {J.~T.}\ \bibnamefont
  {Lange}}, \bibinfo {author} {\bibfnamefont {A.}~\bibnamefont {P{\'e}rez~de
  Alba~Ort{\'i}z}},\ and\ \bibinfo {author} {\bibfnamefont {M.}~\bibnamefont
  {Dijkstra}},\ }\bibfield  {title} {\enquote {\bibinfo {title} {Inverse design
  of crystals and quasicrystals in a non-additive binary mixture of hard
  disks},}\ }\href {https://doi.org/10.1063/5.0210034} {\bibfield  {journal}
  {\bibinfo  {journal} {J. Chem. Phys.}\ }\textbf {\bibinfo {volume} {160}},\
  \bibinfo {pages} {244902} (\bibinfo {year} {2024})}\BibitemShut {NoStop}%
\bibitem [{\citenamefont {Dasgupta}, \citenamefont {Coli},\ and\ \citenamefont
  {Dijkstra}(2020)}]{Dasgupta2020}%
  \BibitemOpen
  \bibfield  {author} {\bibinfo {author} {\bibfnamefont {T.}~\bibnamefont
  {Dasgupta}}, \bibinfo {author} {\bibfnamefont {G.~M.}\ \bibnamefont {Coli}},\
  and\ \bibinfo {author} {\bibfnamefont {M.}~\bibnamefont {Dijkstra}},\
  }\bibfield  {title} {\enquote {\bibinfo {title} {Tuning the glass transition:
  Enhanced crystallization of the {Laves} phases in nearly hard spheres},}\
  }\href {https://doi.org/10.1021/acsnano.9b07090} {\bibfield  {journal}
  {\bibinfo  {journal} {ACS Nano}\ }\textbf {\bibinfo {volume} {14}},\ \bibinfo
  {pages} {3957--3968} (\bibinfo {year} {2020})}\BibitemShut {NoStop}%
\bibitem [{\citenamefont {Schmiedeberg}\ \emph {et~al.}(2011)\citenamefont
  {Schmiedeberg}, \citenamefont {Haxton}, \citenamefont {Nagel},\ and\
  \citenamefont {Liu}}]{Schmiedeberg2011}%
  \BibitemOpen
  \bibfield  {author} {\bibinfo {author} {\bibfnamefont {M.}~\bibnamefont
  {Schmiedeberg}}, \bibinfo {author} {\bibfnamefont {T.~K.}\ \bibnamefont
  {Haxton}}, \bibinfo {author} {\bibfnamefont {S.~R.}\ \bibnamefont {Nagel}},\
  and\ \bibinfo {author} {\bibfnamefont {A.~J.}\ \bibnamefont {Liu}},\
  }\bibfield  {title} {\enquote {\bibinfo {title} {Mapping the glassy dynamics
  of soft spheres onto hard-sphere behavior},}\ }\href
  {https://doi.org/10.1209/0295-5075/96/36010} {\bibfield  {journal} {\bibinfo
  {journal} {Europhys. Lett.}\ }\textbf {\bibinfo {volume} {96}},\ \bibinfo
  {pages} {36010} (\bibinfo {year} {2011})}\BibitemShut {NoStop}%
\bibitem [{\citenamefont {Dyre}(2016)}]{Dyre2016}%
  \BibitemOpen
  \bibfield  {author} {\bibinfo {author} {\bibfnamefont {J.~C.}\ \bibnamefont
  {Dyre}},\ }\bibfield  {title} {\enquote {\bibinfo {title} {Simple liquids'
  quasiuniversality and the hard-sphere paradigm},}\ }\href
  {https://doi.org/10.1088/0953-8984/28/32/323001} {\bibfield  {journal}
  {\bibinfo  {journal} {J. Phys.: Condens. Matter}\ }\textbf {\bibinfo {volume}
  {28}},\ \bibinfo {pages} {323001} (\bibinfo {year} {2016})}\BibitemShut
  {NoStop}%
\bibitem [{\citenamefont {Alkemade}\ \emph {et~al.}(2026)\citenamefont
  {Alkemade}, \citenamefont {Salo}, \citenamefont {Filion},\ and\ \citenamefont
  {Smallenburg}}]{Alkemade2026}%
  \BibitemOpen
  \bibfield  {author} {\bibinfo {author} {\bibfnamefont {R.~M.}\ \bibnamefont
  {Alkemade}}, \bibinfo {author} {\bibfnamefont {A.}~\bibnamefont {Salo}},
  \bibinfo {author} {\bibfnamefont {L.}~\bibnamefont {Filion}},\ and\ \bibinfo
  {author} {\bibfnamefont {F.}~\bibnamefont {Smallenburg}},\ }\bibfield
  {title} {\enquote {\bibinfo {title} {Determining fluid--crystal phase
  boundaries for a binary hard-sphere mixture using direct-coexistence
  simulations},}\ }\href {https://doi.org/10.1063/5.0321591} {\bibfield
  {journal} {\bibinfo  {journal} {J. Chem. Phys.}\ }\textbf {\bibinfo {volume}
  {164}},\ \bibinfo {pages} {124104} (\bibinfo {year} {2026})}\BibitemShut
  {NoStop}%
\bibitem [{\citenamefont {Thompson}\ \emph {et~al.}(2022)\citenamefont
  {Thompson}, \citenamefont {Aktulga}, \citenamefont {Berger}, \citenamefont
  {Bolintineanu}, \citenamefont {Brown}, \citenamefont {Crozier}, \citenamefont
  {{in 't Veld}}, \citenamefont {Kohlmeyer}, \citenamefont {Moore},
  \citenamefont {Nguyen}, \citenamefont {Shan}, \citenamefont {Stevens},
  \citenamefont {Tranchida}, \citenamefont {Trott},\ and\ \citenamefont
  {Plimpton}}]{Thompson2022}%
  \BibitemOpen
  \bibfield  {author} {\bibinfo {author} {\bibfnamefont {A.~P.}\ \bibnamefont
  {Thompson}}, \bibinfo {author} {\bibfnamefont {H.~M.}\ \bibnamefont
  {Aktulga}}, \bibinfo {author} {\bibfnamefont {R.}~\bibnamefont {Berger}},
  \bibinfo {author} {\bibfnamefont {D.~S.}\ \bibnamefont {Bolintineanu}},
  \bibinfo {author} {\bibfnamefont {W.~M.}\ \bibnamefont {Brown}}, \bibinfo
  {author} {\bibfnamefont {P.~S.}\ \bibnamefont {Crozier}}, \bibinfo {author}
  {\bibfnamefont {P.~J.}\ \bibnamefont {{in 't Veld}}}, \bibinfo {author}
  {\bibfnamefont {A.}~\bibnamefont {Kohlmeyer}}, \bibinfo {author}
  {\bibfnamefont {S.~G.}\ \bibnamefont {Moore}}, \bibinfo {author}
  {\bibfnamefont {T.~D.}\ \bibnamefont {Nguyen}}, \bibinfo {author}
  {\bibfnamefont {R.}~\bibnamefont {Shan}}, \bibinfo {author} {\bibfnamefont
  {M.~J.}\ \bibnamefont {Stevens}}, \bibinfo {author} {\bibfnamefont
  {J.}~\bibnamefont {Tranchida}}, \bibinfo {author} {\bibfnamefont
  {C.}~\bibnamefont {Trott}},\ and\ \bibinfo {author} {\bibfnamefont {S.~J.}\
  \bibnamefont {Plimpton}},\ }\bibfield  {title} {\enquote {\bibinfo {title}
  {{LAMMPS} -- a flexible simulation tool for particle-based materials modeling
  at the atomic, meso, and continuum scales},}\ }\href
  {https://doi.org/10.1016/j.cpc.2021.108171} {\bibfield  {journal} {\bibinfo
  {journal} {Comput. Phys. Commun.}\ }\textbf {\bibinfo {volume} {271}},\
  \bibinfo {pages} {108171} (\bibinfo {year} {2022})}\BibitemShut {NoStop}%
\bibitem [{\citenamefont {D{\'\i}az-Pozuelo}, \citenamefont {Lomba},\ and\
  \citenamefont {Gonz{\'a}lez~Noya}(2026)}]{MCCheckerboard2026}%
  \BibitemOpen
  \bibfield  {author} {\bibinfo {author} {\bibfnamefont {A.}~\bibnamefont
  {D{\'\i}az-Pozuelo}}, \bibinfo {author} {\bibfnamefont {E.}~\bibnamefont
  {Lomba}},\ and\ \bibinfo {author} {\bibfnamefont {E.}~\bibnamefont
  {Gonz{\'a}lez~Noya}},\ }\href@noop {} {\enquote {\bibinfo {title}
  {{MC\_Checkerboard}: {GPU}-accelerated {M}onte {C}arlo for patchy particles
  and simple fluid mixtures},}\ }\bibinfo {howpublished}
  {\url{https://github.com/adpozuelo/MC_Checkerboard}} (\bibinfo {year}
  {2026})\BibitemShut {NoStop}%
\bibitem [{\citenamefont {Lomba}, \citenamefont {Diaz-Pozuelo},\ and\
  \citenamefont {Lomba-Moreno}(2026)}]{trjAnalysis2026}%
  \BibitemOpen
  \bibfield  {author} {\bibinfo {author} {\bibfnamefont {E.}~\bibnamefont
  {Lomba}}, \bibinfo {author} {\bibfnamefont {A.}~\bibnamefont
  {Diaz-Pozuelo}},\ and\ \bibinfo {author} {\bibfnamefont {R.}~\bibnamefont
  {Lomba-Moreno}},\ }\href@noop {} {\enquote {\bibinfo {title}
  {trj\_analysis},}\ }\bibinfo {howpublished}
  {\url{https://github.com/elomba/trj_analysis}} (\bibinfo {year} {2026}),\
  \bibinfo {note} {preprint: SSRN, doi:10.2139/ssrn.6875026}\BibitemShut
  {NoStop}%
\bibitem [{\citenamefont {Hansen}\ and\ \citenamefont
  {McDonald}(2013)}]{Hansen2013}%
  \BibitemOpen
  \bibfield  {author} {\bibinfo {author} {\bibfnamefont {J.-P.}\ \bibnamefont
  {Hansen}}\ and\ \bibinfo {author} {\bibfnamefont {I.~R.}\ \bibnamefont
  {McDonald}},\ }\href@noop {} {\emph {\bibinfo {title} {Theory of Simple
  Liquids: With Applications to Soft Matter}}},\ \bibinfo {edition} {4th}\ ed.\
  (\bibinfo  {publisher} {Academic Press},\ \bibinfo {year} {2013})\BibitemShut
  {NoStop}%
\bibitem [{\citenamefont {G{\"o}tze}(2009)}]{Gotze2009}%
  \BibitemOpen
  \bibfield  {author} {\bibinfo {author} {\bibfnamefont {W.}~\bibnamefont
  {G{\"o}tze}},\ }\href@noop {} {\emph {\bibinfo {title} {Complex Dynamics of
  Glass-Forming Liquids: A Mode-Coupling Theory}}}\ (\bibinfo  {publisher}
  {Oxford University Press},\ \bibinfo {year} {2009})\BibitemShut {NoStop}%
\bibitem [{\citenamefont {Steinhardt}, \citenamefont {Nelson},\ and\
  \citenamefont {Ronchetti}(1983)}]{Steinhardt1983}%
  \BibitemOpen
  \bibfield  {author} {\bibinfo {author} {\bibfnamefont {P.~J.}\ \bibnamefont
  {Steinhardt}}, \bibinfo {author} {\bibfnamefont {D.~R.}\ \bibnamefont
  {Nelson}},\ and\ \bibinfo {author} {\bibfnamefont {M.}~\bibnamefont
  {Ronchetti}},\ }\bibfield  {title} {\enquote {\bibinfo {title}
  {Bond-orientational order in liquids and glasses},}\ }\href
  {https://doi.org/10.1103/PhysRevB.28.784} {\bibfield  {journal} {\bibinfo
  {journal} {Phys. Rev. B}\ }\textbf {\bibinfo {volume} {28}},\ \bibinfo
  {pages} {784--805} (\bibinfo {year} {1983})}\BibitemShut {NoStop}%
\bibitem [{\citenamefont {ten Wolde}, \citenamefont {Ruiz-Montero},\ and\
  \citenamefont {Frenkel}(1996)}]{tenWolde1996}%
  \BibitemOpen
  \bibfield  {author} {\bibinfo {author} {\bibfnamefont {P.~R.}\ \bibnamefont
  {ten Wolde}}, \bibinfo {author} {\bibfnamefont {M.~J.}\ \bibnamefont
  {Ruiz-Montero}},\ and\ \bibinfo {author} {\bibfnamefont {D.}~\bibnamefont
  {Frenkel}},\ }\bibfield  {title} {\enquote {\bibinfo {title} {Numerical
  calculation of the rate of crystal nucleation in a {L}ennard-{J}ones system
  at moderate undercooling},}\ }\href {https://doi.org/10.1063/1.471721}
  {\bibfield  {journal} {\bibinfo  {journal} {J. Chem. Phys.}\ }\textbf
  {\bibinfo {volume} {104}},\ \bibinfo {pages} {9932--9947} (\bibinfo {year}
  {1996})}\BibitemShut {NoStop}%
\bibitem [{\citenamefont {Wendt}\ and\ \citenamefont
  {Abraham}(1978)}]{Wendt1978}%
  \BibitemOpen
  \bibfield  {author} {\bibinfo {author} {\bibfnamefont {H.~R.}\ \bibnamefont
  {Wendt}}\ and\ \bibinfo {author} {\bibfnamefont {F.~F.}\ \bibnamefont
  {Abraham}},\ }\bibfield  {title} {\enquote {\bibinfo {title} {Empirical
  criterion for the glass transition region based on {M}onte {C}arlo
  simulations},}\ }\href {https://doi.org/10.1103/PhysRevLett.41.1244}
  {\bibfield  {journal} {\bibinfo  {journal} {Phys. Rev. Lett.}\ }\textbf
  {\bibinfo {volume} {41}},\ \bibinfo {pages} {1244--1246} (\bibinfo {year}
  {1978})}\BibitemShut {NoStop}%
\bibitem [{\citenamefont {Wang}(2012)}]{Wang2012}%
  \BibitemOpen
  \bibfield  {author} {\bibinfo {author} {\bibfnamefont {W.~H.}\ \bibnamefont
  {Wang}},\ }\bibfield  {title} {\enquote {\bibinfo {title} {The elastic
  properties, elastic models and elastic perspectives of metallic glasses},}\
  }\href {https://doi.org/10.1016/j.pmatsci.2011.07.001} {\bibfield  {journal}
  {\bibinfo  {journal} {Prog. Mater. Sci.}\ }\textbf {\bibinfo {volume} {57}},\
  \bibinfo {pages} {487--656} (\bibinfo {year} {2012})}\BibitemShut {NoStop}%
\bibitem [{\citenamefont {Lewandowski}, \citenamefont {Wang},\ and\
  \citenamefont {Greer}(2005)}]{Lewandowski2005}%
  \BibitemOpen
  \bibfield  {author} {\bibinfo {author} {\bibfnamefont {J.~J.}\ \bibnamefont
  {Lewandowski}}, \bibinfo {author} {\bibfnamefont {W.~H.}\ \bibnamefont
  {Wang}},\ and\ \bibinfo {author} {\bibfnamefont {A.~L.}\ \bibnamefont
  {Greer}},\ }\bibfield  {title} {\enquote {\bibinfo {title} {Intrinsic
  plasticity or brittleness of metallic glasses},}\ }\href
  {https://doi.org/10.1080/09500830500080474} {\bibfield  {journal} {\bibinfo
  {journal} {Philos. Mag. Lett.}\ }\textbf {\bibinfo {volume} {85}},\ \bibinfo
  {pages} {77--87} (\bibinfo {year} {2005})}\BibitemShut {NoStop}%
\end{thebibliography}%
\end{document}